\documentclass[%
 aip,
 amsmath,amssymb,
preprint,%
]{revtex4-1}

\usepackage{graphicx}% Include figure files
\usepackage{dcolumn}% Align table columns on decimal point
\usepackage{bm}% bold math
\usepackage[utf8]{inputenc}
\usepackage[T1]{fontenc}
\usepackage{mathptmx}
\usepackage{etoolbox}

\makeatletter
\def\@email#1#2{%
 \endgroup
 \patchcmd{\titleblock@produce}
  {\frontmatter@RRAPformat}
  {\frontmatter@RRAPformat{\produce@RRAP{*#1\href{mailto:#2}{#2}}}\frontmatter@RRAPformat}
  {}{}
}%
\makeatother

\graphicspath{{./figure}}

\begin{document}

%\preprint{AIP/123-QED}

\title[Numerical analysis of ligament instability and breakup in shear flow]
{Numerical analysis of ligament instability and breakup in shear flow}
\author{Hideki Yanaoka}
 \altaffiliation[]{}%Lines break automatically or can be forced with \\
 \email{yanaoka@iwate-u.ac.jp}
\affiliation{
Department of Systems Innovation Engineering, 
Faculty of Science and Engineering, Iwate University, 
4-3-5 Ueda, Morioka, Iwate 020-8551, Japan
}%

\author{Kosuke Nakayama}
\affiliation{
DOWA Technology Co., Ltd.., 
22F, Akihabara UDX, 4-14-1 Sotokanda, Chiyoda-ku, Tokyo 101-0021, Japan
}

\date{\today}% It is always \today, today,
             %  but any date may be explicitly specified

\begin{abstract}
In this study, we perform a numerical analysis of the instability 
of a ligament in shear flow 
and investigate the effects of air-liquid shear on the growth rate 
of the ligament interface, breakup time, and droplet diameter 
formed by the breakup. 
The ligament is stretched in the flow direction by the shearing of airflow. 
Furthermore, as the influence of the shear flow increases, 
the ligament becomes deformed into a liquid sheet, 
and a perforation forms at the center of the liquid sheet. 
The liquid sheet breaks up due to the growth of the perforation 
and contracts under the influence of surface tension, 
forming two ligaments with diameters smaller than that of the original ligament. 
The shearing of the airflow causes the original ligament to elongate, 
and the cross-section of the ligament becomes elliptical, 
which increases instability. 
As a result, the growth rate of the ligament exceeds 
the theoretical value, 
and increases with increasing wavenumber of the initial disturbance. 
Therefore, the diameter of the formed droplets in shear flow 
decreases due to the increase in the wavenumber 
that governs the breakup of the ligament, 
and because the growth rate increases, the breakup time for the ligament 
decreases. 
As the velocity difference of the shear flow increases, 
constrictions of the ligament form earlier 
and the diameter of the satellite droplet increases. 
As the diameter of the satellite droplet increases 
and that of the main droplet decreases, 
the dispersion of the droplet diameter decreases, 
making the diameter uniform.
\end{abstract}

\maketitle

%\begin{quotation}
%\end{quotation}

%##############################################################################
\section{Introduction}
%##############################################################################

Liquid atomization technology is used in various fields, 
including industry, agriculture, and medicine. 
In industry, atomization technology is applied in internal combustion engines, spray coating, 
spray drying, and powder production. 
Elucidation of atomization characteristics would thus help improve performance. 
The process of liquid atomization is not fully understood because it consists of very complicated and small-scale phenomena. 
Studies on 
the deformation and breakup of simple-shaped liquids such as liquid sheets, 
liquid columns, and droplets have been conducted to elucidate atomization.

In the general atomization process, 
as the ejected liquid (in the form of a liquid sheet 
or a liquid column) is destabilized by the shearing of gas and liquid, 
a fine-scale liquid column, called a ligament, is generated 
and the liquid eventually splits into small droplets \cite{Hashimoto_1995}. 
Theoretical studies \cite{Senecal_et_al_1999, O'Rourke&Amsden_1987}, 
experimental studies \cite{Yoshida_2000, Negeed_et_al_2011}, and 
numerical studies \cite{Li&Tankin_1991, Lozano_et_al_1998, Li_et_al_2000, Kan&Yoshinaga_2007} 
have been conducted on the destabilization of liquid sheets and droplets. 
Although many studies on the destabilization of liquid columns 
and cylindrical liquid jets have been reported 
\cite{Rayleigh_1878, Weber_1931, Donnelly_et_al_1966, Yuen_1968, Goedde&Yuen_1970, Rutland&Jameson_1971, Lafrance_1975, Sterling&Sleicher_1975, Hoyt&Taylor_1977, Tjahjadi_et_al_1992, Park_et_al_2006, Kasyap_et_al_2009, Hoeve_et_al_2010, Lakdawala_et_al_2015, Wang&Fang_2015}, 
few studies have been conducted on fine-scale ligaments, 
and thus the destabilization of ligaments is not fully understood. 

In previous studies \cite{Fraser_et_al_1962, Dombrowski&Johns_1963}, 
the instability of a ligament formed in the atomization process 
is often considered based on the linear theory 
of the instability of a stationary liquid column \cite{Rayleigh_1878}. 
Rayleigh's theory does not consider the effects of viscosity 
and surrounding fluids. 
The instability of a liquid column in consideration of viscosity 
\cite{Weber_1931, Sterling&Sleicher_1975, Lakdawala_et_al_2015} 
and the instability of a cylindrical liquid jet due to the shearing of 
gas and liquid \cite{Hoyt&Taylor_1977} have been investigated.

In a study of cylindrical liquid jets \cite{Hoyt&Taylor_1977}, 
the formation of small-scale liquid columns was observed 
at the interface of the liquid column destabilized by velocity shear. 
The breakup of this fine-scale secondary liquid column (ligament) 
generates droplets. 
When an external force such as velocity shear acts on a liquid 
that is not limited to a liquid column (i.e., it is a continuum), 
turbulence occurs inside the liquid. 
The growth of this turbulence deforms the liquid interface, leading to the formation of fine ligaments \cite{Hashimoto_1995, Sallam_el_al_1999}. 
Finally, fine sprays form due to the repeated breakup of ligaments 
and droplets. 
Therefore, the breakup characteristics 
of ligaments should be investigated to clarify spray characteristics.

In the high-speed flow field in an atomizer, 
because both the gas and liquid flows are turbulent, 
the turbulence intensity at the gas-liquid interface increases with time. 
In such a flow field, the airflow around a ligament is not uniform; 
instead, it is shear flow with a velocity gradient. 
In this case, because the ligament is deformed
by the shear flow, 
the breakup time and droplet diameter 
will be different from those of a ligament 
in a stationary fluid or a uniform airflow. 
However, the behavior of a ligament in turbulent flow 
and the effect of shear flow on the instability of a ligament 
have not been sufficiently investigated.

To clarify the behavior of a ligament in turbulent flow, 
we perform a numerical analysis of a ligament in shear flow 
and investigate the effects of gas-liquid shear on the growth rate 
at the ligament interface, break time, and droplet size.

%##############################################################################
\section{Numerical Procedures}
%##############################################################################

This study considers incompressible viscous flow. 
We apply the level set method to track the interface 
\cite{Sussman_et_al_1994}. 
In addition, the continuum surface force method \cite{Brackbill_et_al_1992} is used 
to evaluate the surface tension, which 
is treated as the body force. 
The governing equations for incompressible two-phase flow are 
the continuity equation, the Navier-Stokes equation, 
the advection equation for the level set function, 
and the reinitialization equation, which are given as follows:
%------------------------------------------------------------------------------
\begin{equation}
   \nabla \cdot {\bf u} = 0,
   \label{continuity}
\end{equation}
\begin{equation}
   \frac{\partial {\bf u}}{\partial t} + \nabla \cdot ({\bf u} {\bf u}) 
   = \frac{1}{\rho} \left[ - \nabla p 
   + \frac{1}{Re} \nabla \cdot (2 \mu {\bf D}) 
   + \frac{1}{We} \kappa \hat{\bf n} \delta_s \right],
   \label{navier-stokes}
\end{equation}
\begin{equation}
   \frac{\partial \phi}{\partial t} + {\bf u} \cdot \nabla \phi = 0,
   \label{level-set}
\end{equation}
\begin{equation}
   \frac{\partial \phi}{\partial \tau} + {\bf V} \cdot \nabla \phi = S(\phi_0),
   \label{reinitialization}
\end{equation}
\begin{equation}
   {\bf V} = S(\phi_0) \frac{\nabla \phi}{|\nabla \phi|},
\end{equation}
%------------------------------------------------------------------------------
where $t$ is time, ${\bf u}=(u, v, w)$ is the velocity vector 
at the coordinates ${\bf x}=(x, y, z)$, $\rho$ is the density, 
$p$ is the pressure, $\mu$ is the viscosity coefficient, 
and ${\bf D}$ is the strain rate tensor. 
$\kappa$ is the interface curvature, 
$\hat{\bf n}$ is the unit normal vector at the interface, 
$\delta_s$ is the delta function, 
and $\phi$ is the level set function. 
$\phi_0$ is the level set function before reinitialization. 
The variables in the fundamental equations are non-dimensionalized 
as follows using the reference values of length $L_\mathrm{ref}$, 
velocity $U_\mathrm{ref}$, density $\rho_\mathrm{ref}$, 
and viscosity coefficient $\mu_\mathrm{ref}$:
%------------------------------------------------------------------------------
\[
   t^{*} = \frac{U_\mathrm{ref} t}{L_\mathrm{ref}}\ , \quad 
   {\bf x}^{*} = \frac{\bf x}{L_\mathrm{ref}}\ , \quad 
   {\bf u}^{*} = \frac{\bf u}{U_\mathrm{ref}}\ , \quad 
   p^{*} = \frac{p}{\rho U_\mathrm{ref}^{2}}\ , \quad 
   \phi^{*} = \frac{\phi}{L_\mathrm{ref}}, \quad 
\]
\begin{equation}
   \rho^{*} = \frac{\rho}{\rho_\mathrm{ref}}\ , \quad 
   \mu^{*} = \frac{\mu}{\mu_\mathrm{ref}}\ ,
\end{equation}
%------------------------------------------------------------------------------
where the superscript $*$ represents a dimensionless variable 
and is omitted in the above governing equations. 
As dimensionless parameters in these equations, 
$Re=U_\mathrm{ref} L_\mathrm{ref}/\nu$ is the Reynolds number, 
$We=\rho_\mathrm{ref} L_\mathrm{ref} U_\mathrm{ref}^2/\sigma$ 
is the Weber number, and $\sigma$ is the surface tension coefficient. 
The strain rate tensor is defined as
%------------------------------------------------------------------------------
\begin{equation}
   {\bf D} = \frac{1}{2} \left[ \nabla {\bf u} + (\nabla {\bf u})^T \right].
\end{equation}
%------------------------------------------------------------------------------
The density and viscosity coefficients for the gas-liquid two-phase flow 
are expressed as
%------------------------------------------------------------------------------
\begin{equation}
   \rho = \rho_g + (\rho_l - \rho_g) H(\phi), \quad 
   \mu = \mu_g + (\mu_l - \mu_g) H(\phi),
\end{equation}
%------------------------------------------------------------------------------
where the subscripts $l$ and $g$ represent liquid and gas, respectively. 
$H(\phi)$ is the Heaviside function and is given as
%------------------------------------------------------------------------------
\begin{equation}
   H(\phi) = \left\{
   \begin{array}{ll}
      1 & \quad \mbox{if} \quad \phi > \epsilon, \\
      0 & \quad \mbox{if} \quad \phi < -\epsilon, \\
      \displaystyle
      \frac{\phi + \epsilon}{2 \epsilon} 
      + \frac{1}{2 \pi} \sin \left( \frac{\pi \phi}{\epsilon} \right) & 
      \quad \mbox{if} \quad |\phi| \le \epsilon.
   \end{array}
   \right.
\end{equation}
%------------------------------------------------------------------------------
The gas-liquid interface is represented as a transition region 
with a thickness of $2\varepsilon$, where 
$\varepsilon$ is set to $1.0-2.0$ times the grid width. 
The interface curvature and the unit normal vector are expressed as
%------------------------------------------------------------------------------
\begin{equation}
   \kappa = - \nabla \cdot \hat{\bf n},
\end{equation}
\begin{equation}
   \hat{\bf n} = \frac{\nabla \phi}{|\nabla \phi|}.
\end{equation}
%------------------------------------------------------------------------------
The delta function at the interface is obtained as the gradient 
of the Heaviside function as follows:
%------------------------------------------------------------------------------
\begin{equation}
   \delta_\mathrm{s} = \nabla_{\phi} H(\phi) = \left\{
   \begin{array}{ll}
      0 & \quad \mbox{if} \quad |\phi| > \epsilon, \\
      \displaystyle
      \frac{1}{2 \epsilon} \left[ 
      1 + \cos \left( \frac{\pi \phi}{\epsilon} \right) \right] & 
      \quad \mbox{if} \quad |\phi| \le \epsilon.
   \end{array}
   \right.
\end{equation}
%------------------------------------------------------------------------------

The simplified marker and cell method \cite{Amsden&Harlow_1970} is used to solve 
Eqs. (\ref{continuity}) and (\ref{navier-stokes}). 
The Crank-Nicholson method is used to discretize the time derivatives, 
and time marching is performed. 
The second-order central difference scheme is used 
for the discretization of the space derivatives.
For the discretization of Eq. (\ref{level-set}), 
the Crank-Nicholson method is applied for 
the time derivative 
and the fifth-order weighted essentially non-oscillatory (WENO) scheme \cite{Jiang&Peng_2000} is applied 
for the space derivatives.
For Eq. (\ref{reinitialization}), 
the total variation diminishing Runge-Kutta method with third-order accuracy 
\cite{Shu&Osher_1988, Shu&Osher_1989} is applied for the time derivative 
and the fifth-order WENO scheme \cite{Jiang&Peng_2000} is applied 
for the space derivatives.

%##############################################################################
\section{Calculation Conditions}
%##############################################################################

%++++++++++++++++++++++++++++++++++++++++++++++++++++++++++++++++++++++++++++++
\subsection{Comparison with theoretical solution}
\label{subsec1}
%++++++++++++++++++++++++++++++++++++++++++++++++++++++++++++++++++++++++++++++

In this study, we first analyze the instability of the interface 
of a stationary ligament to verify the validity of the above calculation method 
for phenomena in which surface tension has a great effect. 
A cylindrical coordinate system is used for the analysis, 
and an axisymmetric ligament is considered. 
The origin is placed on the central axis of a ligament with diameter $d$. 
The direction of the central axis is the $x$-axis 
and the radial direction is the $r$-axis. 
A disturbance $\eta$, which is a cosine function with a small initial amplitude 
$\eta_0$, is applied to the surface of the ligament with radius $a$.
%------------------------------------------------------------------------------
\begin{equation}
   \eta = \eta_0 \cos (2 \pi x / \lambda),
\end{equation}
%------------------------------------------------------------------------------
where $k=2\pi/\lambda$ is the wavenumber and $\lambda$ is the wavelength. 
The interface of the stationary ligament fluctuates 
due to the influence of this disturbance. 
This study varies the wavenumber $k$ of the initial disturbance 
and compares the growth rate of the interface for each wavenumber 
with the theoretical value for the liquid column 
\cite{Rayleigh_1878, Weber_1931}. 
We perform axisymmetric two-dimensional analyses 
for inviscid and viscous fluids.

The calculation region is $\lambda$ in the axial direction 
and $2a$ in the radial direction. 
Three uniform grids, with dimensions of $81 \times 26$ (grid1), $101 \times 51$ (grid2), 
and $121 \times 101$ (grid3) are used for the calculations. 
The minimum grid widths are $0.04d$, $0.02d$, and $0.01d$, respectively. 
As described later, 
in the analyses for inviscid and viscous fluids, 
we confirmed that appropriate results are obtained 
even with the dimensions of grid2 and grid1.

Regarding the initial conditions in this calculation, the fluid is stationary, 
and a small disturbance with an initial amplitude of $\eta_0$ is applied to the interface. 
Periodic boundary conditions for velocity are applied at $x=0$ and $\lambda$, 
symmetry conditions are assumed at $r=0$, 
and slip boundary conditions are applied at $r=2a$. 
For the level set function, periodic boundary conditions are applied 
at $x=0$ and $\lambda$, symmetry conditions are assumed at $r=0$, 
and the boundary value at $r=2a$ is obtained by extrapolation.

In our calculation, the diameter of the ligament is $d=1.0 \times 10^{-5}$ m. 
The liquid is a light oil and the surrounding gas is air. 
The densities of the liquid and gas are $\rho_l=851$ kg/m$^3$ 
and $\rho_g=1.20$ kg/m$^3$ 
and the viscosity coefficients are $\mu_l=2.225 \times 10^{-3}$ Pa s 
and $\mu_g=1.827 \times 10^{-5}$ Pa s, respectively. 
The surface tension coefficient is $\sigma=2.44 \times 10^{-2}$ N/m. 
The reference values used in the calculation are $L_\mathrm{ref}=d$ 
and $U_\mathrm{ref}=d [\sigma/(\rho_l d^3)]^{1/2}$. 
The dimensionless time $T$ is $T=t/(\rho_l d^3/\sigma)^{1/2}$ 
and the Weber number $We$ is $We=1.0$. 
In the calculation considering viscosity, 
the Ohnesorge number, which is defined as 
$Oh=\mu_l/(\rho_l \sigma L_\mathrm{ref})^{1/2}$, 
is set to $Oh=0.15$. 
Under these conditions, the Reynolds number is $Re=6.67$. 
The time intervals used in the calculation are $\Delta t/(d/U_\mathrm{ref})=8.0 \times 10^{-4}$, $4.0 \times 10^{-4}$, and $2.0 \times 10^{-4}$ 
for grid1, grid2, and grid3, respectively. 
Calculations are performed for dimensionless wavenumbers 
$ka=0.3$, 0.5, 0.7, and 0.9, 
and the growth rate of the ligament interface at each wavenumber 
is compared with the theoretical results for a liquid column 
\cite{Rayleigh_1878, Weber_1931}. 
The growth rate $\omega_0$ of the interface for the inviscid liquid column 
is defined by linear theory as
%------------------------------------------------------------------------------
\begin{equation}
   \omega_{0}^2 = \frac{\sigma}{\rho_{l} a^3} 
                  \frac{\alpha I_1(\alpha)}{I_0(\alpha)} (1 - \alpha^2),
\end{equation}
%------------------------------------------------------------------------------
where $\alpha=ka$ and $In(\alpha)$ is a modified Bessel function of the first kind. 
The dimensionless growth rate $\Omega_0$ is defined as 
$\Omega_0=\omega_0/[\sigma/(\rho_l a^3)]^{1/2}$. 
From linear theory, the growth rate $\omega$ of the interface for the viscous liquid column 
is defined as
%------------------------------------------------------------------------------
\begin{equation}
   \omega^2 + \frac{3 \mu_{l} \alpha^2}{\rho_{l} a^2} \omega 
            = \frac{\sigma}{2 \rho_{l} a^3}(1 - \alpha^2) \alpha^2.
\end{equation}
%------------------------------------------------------------------------------
The dimensionless growth rate $\Omega$ is defined as 
$\Omega=\omega/[\sigma/(\rho_l a^3)]^{1/2}$.

To examine the effect of the initial amplitude on the growth of the interface, 
the results obtained using three amplitudes ($\eta_0/d=0.01$, 0.001, 
and 0.0001) were compared. 
There was no difference in the results for $\eta_0/d \le 0.01$. 
It was found that $\eta_0/d=0.01$ was sufficiently small. 
Therefore, the results obtained using $\eta_0/d=0.01$ are shown below.

%++++++++++++++++++++++++++++++++++++++++++++++++++++++++++++++++++++++++++++++
\subsection{Instability analysis of ligament in shear flow}
%++++++++++++++++++++++++++++++++++++++++++++++++++++++++++++++++++++++++++++++

%------------------------------------------------------------------------------
% Figure 1
%------------------------------------------------------------------------------
\begin{figure}[!t]
\includegraphics[width=70mm]{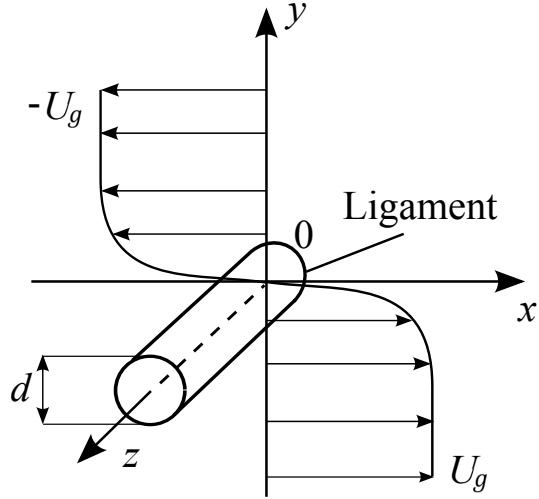} \\
\vspace*{-0.5\baselineskip}
\caption{Flow configuration and coordinate system for ligament in shear flow}
\label{flow_model3d}
\end{figure}
%------------------------------------------------------------------------------
Next, we investigate the instability of a ligament in a shear flow. 
Figure \ref{flow_model3d} shows the flow configuration and coordinate system for this ligament. 
The origin is placed on the central axis of a ligament with diameter $d$. 
The horizontal direction is the $x$-axis, 
the vertical direction is the $y$-axis, 
and the central axis of the ligament is the $z$-axis. 
The gas moves in the positive and negative directions of the $x$-axis 
at a velocity $U_g$ on the lower and upper sides of the ligament, 
respectively. 
The velocity difference of the shear flow is $\Delta U=2U_g$. 
An initial disturbance with wavelength $\lambda$ is applied to
the surface of the ligament.

The calculation region is $10d$ in the $x$-axis direction, 
$8d$ in the $y$-axis direction, and $\lambda$ in the $z$-axis direction. 
In the analysis described in subsection \ref{subsec1}, 
appropriate results were obtained even with the minimum grid width of grid1 
in the two-dimensional viscous analysis. 
Thus, in this three-dimensional viscous analysis, 
we use a non-uniform grid with dimensions of $159 \times 141 \times 81$ (grid2) 
and a minimum grid width of $0.04d$. 
Using grids with dimensions of $103 \times 87 \times 41$ (grid1) 
and $209 \times 197 \times 101$ (grid3) 
and minimum grid widths of $0.08d$ and $0.02d$, respectively, 
we analyzed the behavior of the interface in the shear flow 
and confirmed that there is no grid dependence on the interface 
and velocity distributions. 
The droplet diameter after ligament breakup is compared with 
the results of previously reported experiments and calculations 
to show the validity of our calculation results.

As the initial condition in this calculation, 
the following velocity is applied to the gas:
%------------------------------------------------------------------------------
\begin{equation}
    u = -U_g \tanh (y / \delta),
\end{equation}
%------------------------------------------------------------------------------
where $\delta$ is a parameter related to the velocity boundary layer thickness. 
In this study, we set $\delta/d=0.5$ 
and set the velocity boundary layer thickness to about the radius of 
the ligament so that the entire ligament is in an airflow 
with a velocity gradient.

This study defines the dimensionless velocity difference as 
$\Delta U^*=\Delta U/U_\mathrm{ref}$ by non-dimensionalizing 
the velocity difference $\Delta U$ using the reference velocity 
$U_\mathrm{ref}$, as described later. 
The gas velocity difference $\Delta U$ in this study is determined 
by referring to previous diesel spray experiments 
\cite{Arai&Hiroyasu_1993, Suh_et_al_2007, Suh&Lee_2008, Kim&Lee_2008, Zama_et_al_2012}. 
The velocity of droplets in diesel spray depends on the injection pressure 
and atmospheric density. 
It is approximately 0-100 m/s for a low-speed spray and 0-200 m/s 
for a high-speed spray. 
In this study, we set $\Delta U=0$, 10, 20, 40, 60, 80, and 100 m/s 
for incompressible flow. 
The dimensionless velocity difference is $\Delta U^*=0$, 
5.9, 11.9, 23.7, 35.6, 47.4, and 59.2, respectively.

As in the previous subsection, an initial disturbance with a small amplitude 
$\eta_0$ is applied to the surface of the ligament.
Regarding the boundary conditions for the velocity and level set function, 
periodic boundary conditions are given for the $x$- and $z$-axes. 
At the upper and lower boundaries in the $y$-axis direction, 
uniform flow velocities $-U_g$ and $U_g$ are given, respectively, 
and the level set function is extrapolated.

In this calculation, the liquid is a light oil and the gas is air. 
The physical properties are the same as those described 
in subsection \ref{subsec1}. 
The reference values of length and velocity are defined as 
$L_\mathrm{ref}=d$ and $U_\mathrm{ref}=[\sigma/(\rho_l d^3)]^{1/2}$, respectively,
and the Ohnesorge number is defined as 
$Oh=\mu_l/(\rho_l \sigma L_\mathrm{ref})^{1/2}$.

In our calculation, the ligament diameter is determined 
by back calculation from the droplet diameter using linear theory 
\cite{Rayleigh_1878}. 
The diameter of a droplet formed by the atomizer depends on several factors,
such as the nozzle diameter, injection pressure, and nozzle shape, 
but is generally about $1.0 \times 10^{-5}$ $-2.0 \times 10^{-4}$ m 
\cite{Tamaki_et_al_2001b, Lee&Park_2002, Liu_et_al_2006, Suh_et_al_2007, Suh&Lee_2008, Kim&Lee_2008}. 
In linear theory \cite{Rayleigh_1878}, 
the relationship between the diameter $d$ of the liquid column 
and the diameter $d_\mathrm{drop}$ of the droplet is expressed as 
$d_\mathrm{drop}=1.89d$.
From this, the ligament diameter is 
$d=5.3 \times 10^{-6}$-$1.0 \times 10^{-4}$ m. 
In this calculation, the ligament diameter is set to $d=1.0 \times 10^{-5}$ m. 
Then, the Reynolds number, Weber number, and Ohnesorge number 
are $Re=6.67$, $We=1.0$, and $Oh=0.15$, respectively. 
The time interval used in the calculation is $\Delta t/(d/\Delta U)=0.0005$.

%##############################################################################
\section{Results and Discussion}
%##############################################################################

%++++++++++++++++++++++++++++++++++++++++++++++++++++++++++++++++++++++++++++++
\subsection{Comparison with theoretical solution}
%++++++++++++++++++++++++++++++++++++++++++++++++++++++++++++++++++++++++++++++

\begin{figure}[!t]
\begin{minipage}{0.48\linewidth}
\includegraphics[trim=0mm 0mm 0mm 0mm, clip, width=70mm]{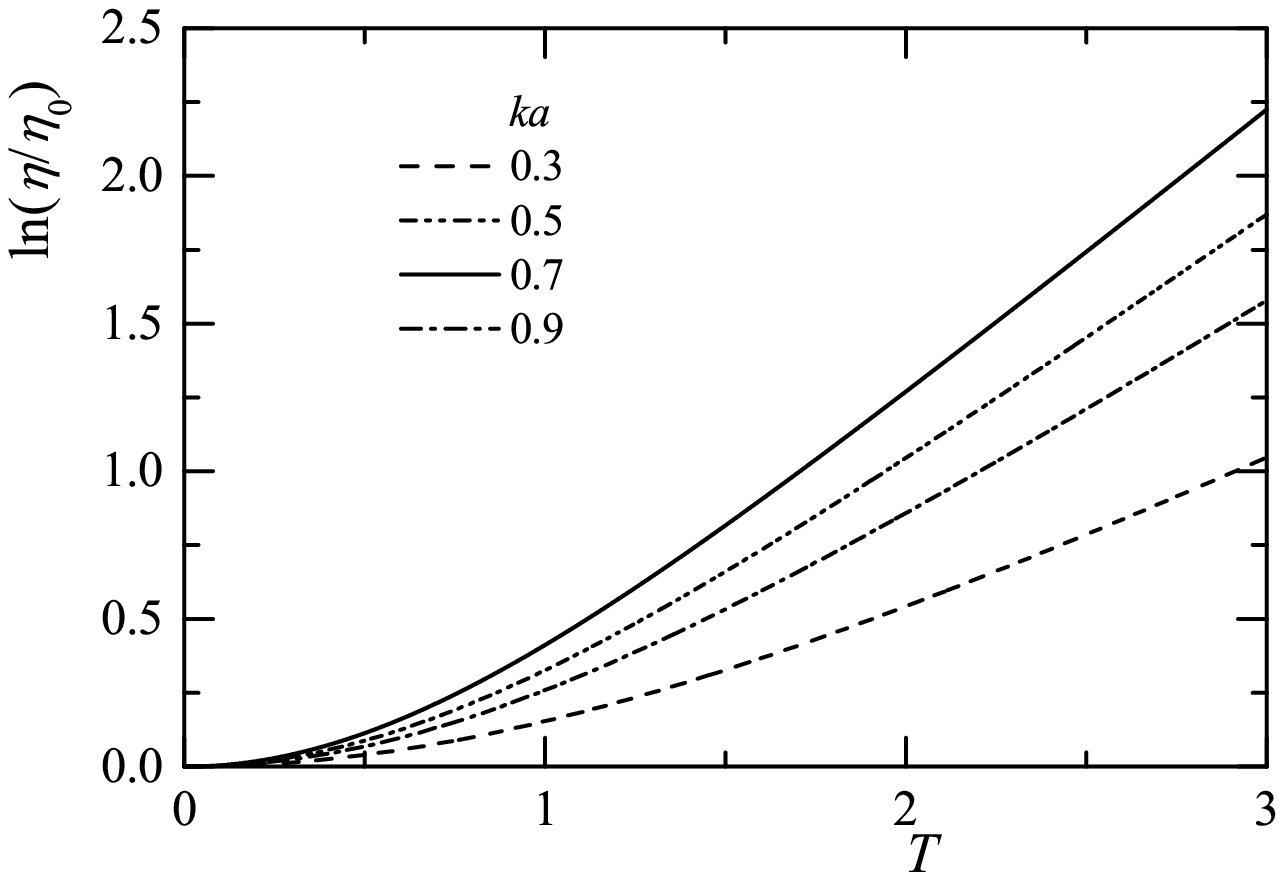} \\
(a) inviscid ligament
\end{minipage}
\hspace{0.02\linewidth}
\begin{minipage}{0.48\linewidth}
\includegraphics[trim=0mm 0mm 0mm 0mm, clip, width=70mm]{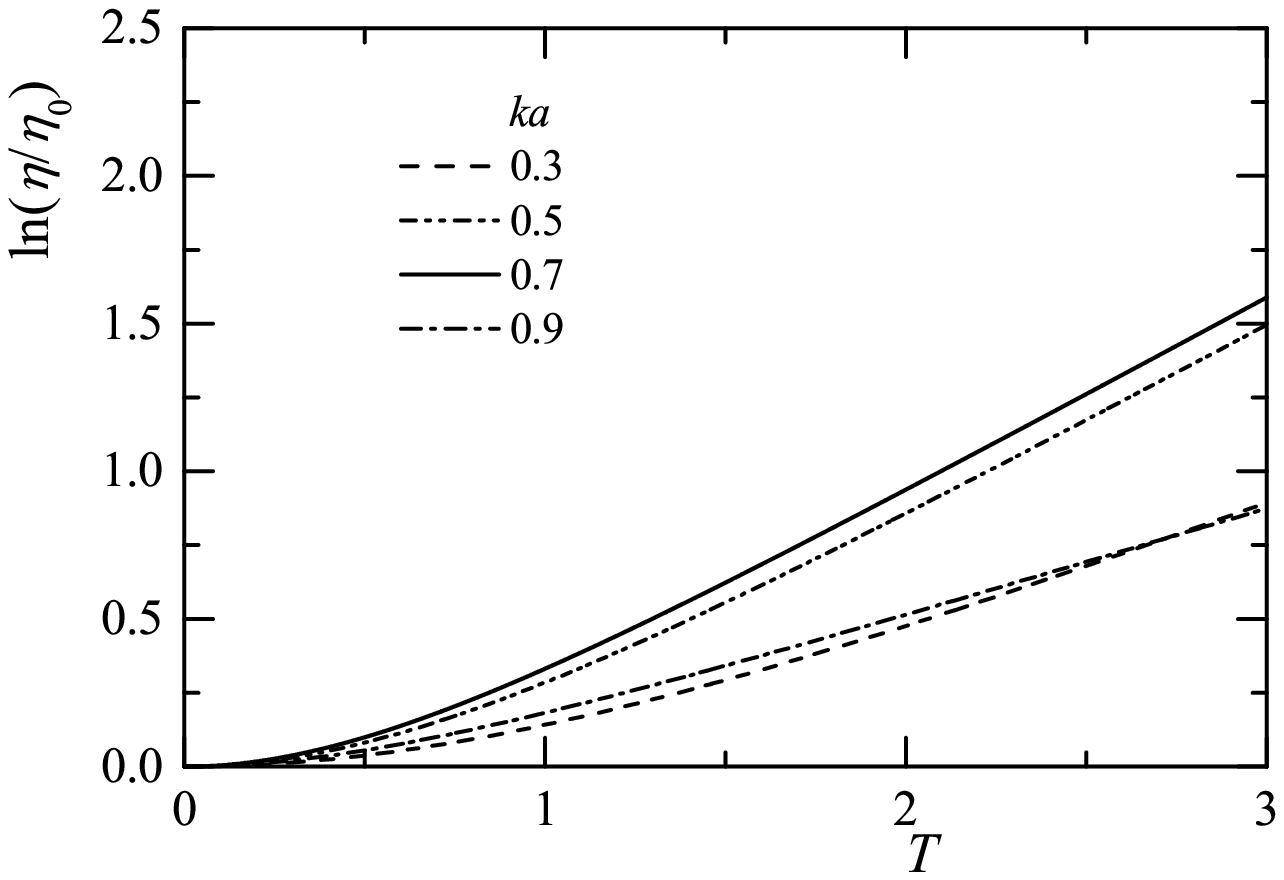} \\
(b) viscous ligament
\end{minipage}
\caption{Time variations of amplitude for stationary ligament}
\label{amp_2d}
\end{figure}

To investigate the growth rate of the ligament interface, 
Fig. \ref{amp_2d} shows the time variation of the amplitude of the interface. 
The results for inviscid and viscous ligaments are shown on the left and right sides, 
respectively. 
The amplitude is the average value of the amplitudes at the center 
and end of the ligament. 
At all wavenumbers, the amplitude of the interface increases exponentially 
with time. 
The growth of the interface amplitude is the fastest at $ka=0.7$ 
and the slowest at $ka=0.3$. 
The growth of the interface of the viscous ligament is slower overall 
than that of the inviscid ligament.

%------------------------------------------------------------------------------
% Figure 2
%------------------------------------------------------------------------------
%------------------------------------------------------------------------------

%------------------------------------------------------------------------------
% Figure 3
%------------------------------------------------------------------------------
\begin{figure}[!t]
\includegraphics[trim=0mm 0mm 0mm 0mm, clip, width=80mm]{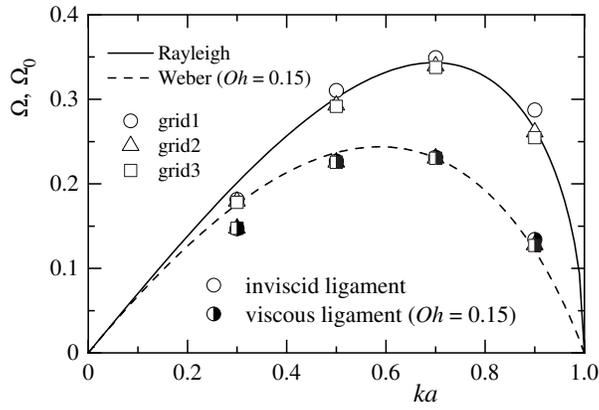}
\vspace*{-0.5\baselineskip}
\caption{Growth rate of interface for stationary ligament}
\label{grate_2d}
\end{figure}
%------------------------------------------------------------------------------
The validity of our calculation method is verified 
by comparing the growth rate of the ligament interface 
with the theoretical values \cite{Rayleigh_1878, Weber_1931}. 
Figure \ref{grate_2d} compares the dimensionless growth rates 
$\Omega_0$ and $\Omega$ obtained using each grid 
with the theoretical values \cite{Rayleigh_1878, Weber_1931}. 
In the analysis for the inviscid ligament, the calculation results using the three grids 
agree well with the theoretical values at all wavenumbers. 
In the analysis for the viscous ligament, the calculation results at all wavenumbers 
agree well with the theoretical values, 
and there is no difference in the results using the three grids. 
Our calculation method captures the behavior of the ligament 
under consideration of viscosity even with the dimensions of grid1. 
These results indicate that the behavior of 
the ligament interface deformed by surface tension was captured in this analysis 
and that grid2 had sufficient resolution.

%++++++++++++++++++++++++++++++++++++++++++++++++++++++++++++++++++++++++++++++
\subsection{Deformation of ligament due to shear flow}
%++++++++++++++++++++++++++++++++++++++++++++++++++++++++++++++++++++++++++++++

%------------------------------------------------------------------------------
% Figure 4
%------------------------------------------------------------------------------
\begin{figure}[!t]
\begin{minipage}{0.48\linewidth}
\includegraphics[trim=0mm 0mm 0mm 0mm, clip, width=70mm]{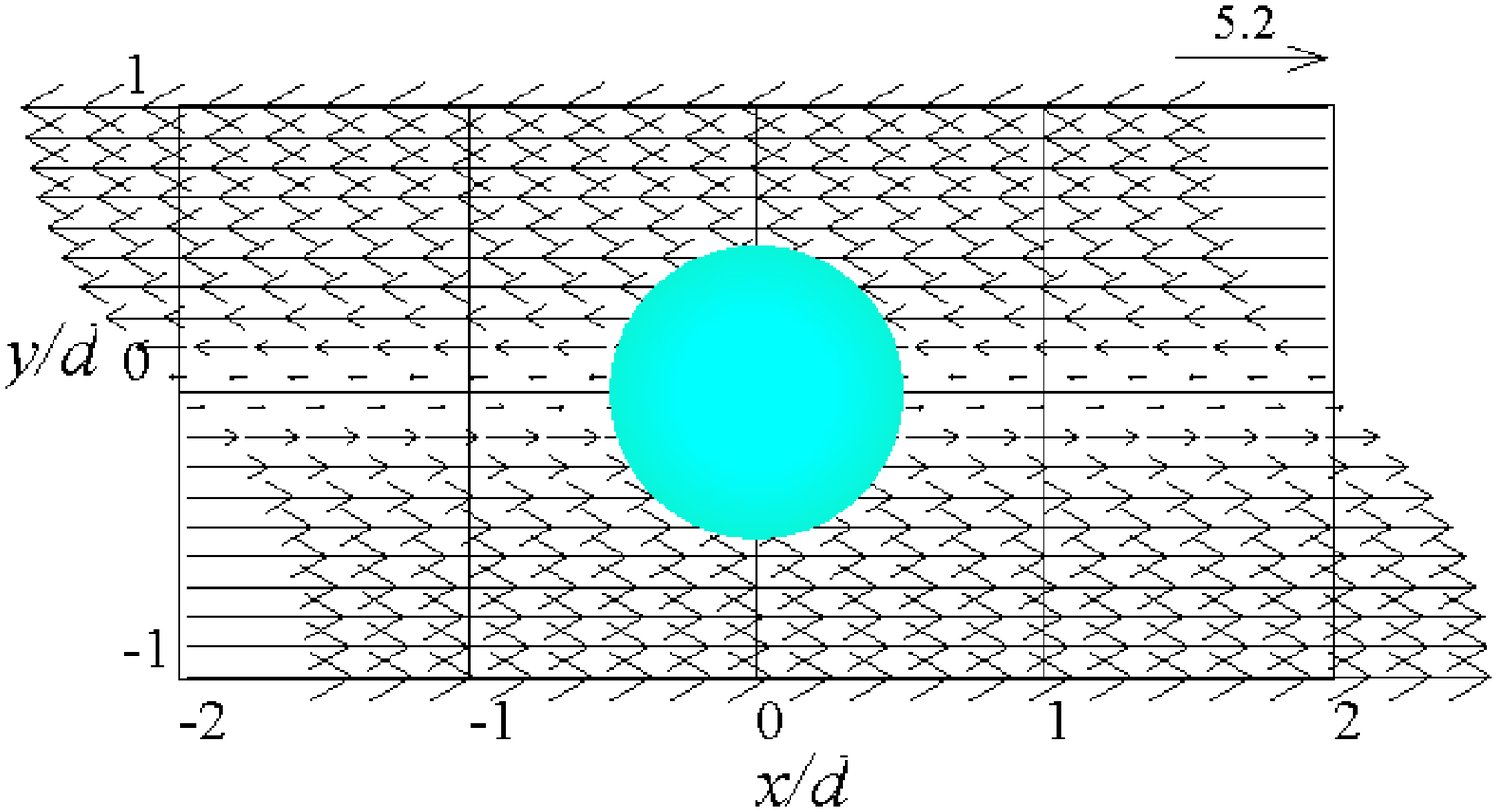} \\
(a) $T=0$
\end{minipage}
\hspace{0.02\linewidth}
\begin{minipage}{0.48\linewidth}
\includegraphics[trim=0mm 0mm 0mm 0mm, clip, width=70mm]{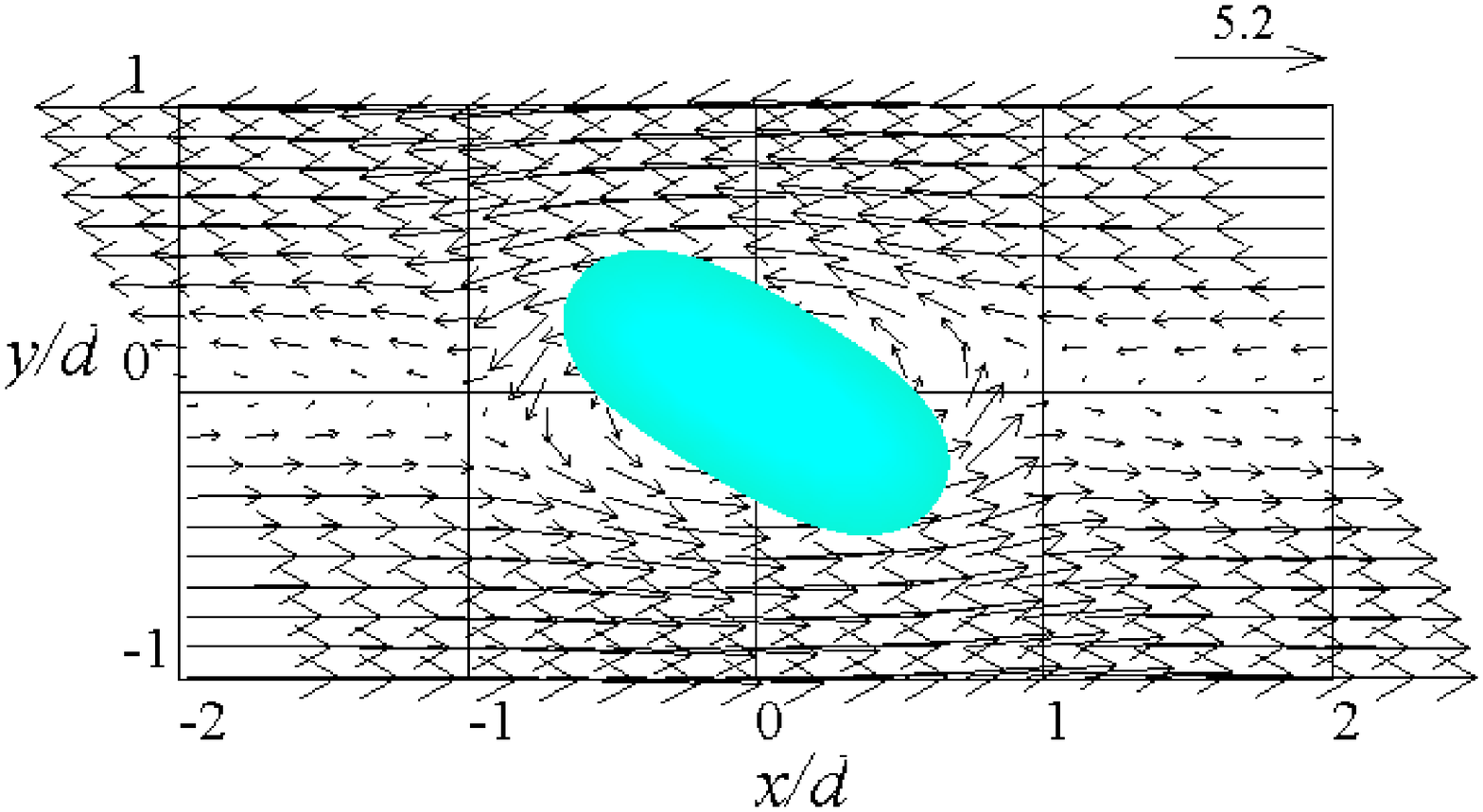} \\
(b) $T=0.1$
\end{minipage}

\begin{minipage}{0.48\linewidth}
\includegraphics[trim=0mm 0mm 0mm 0mm, clip, width=70mm]{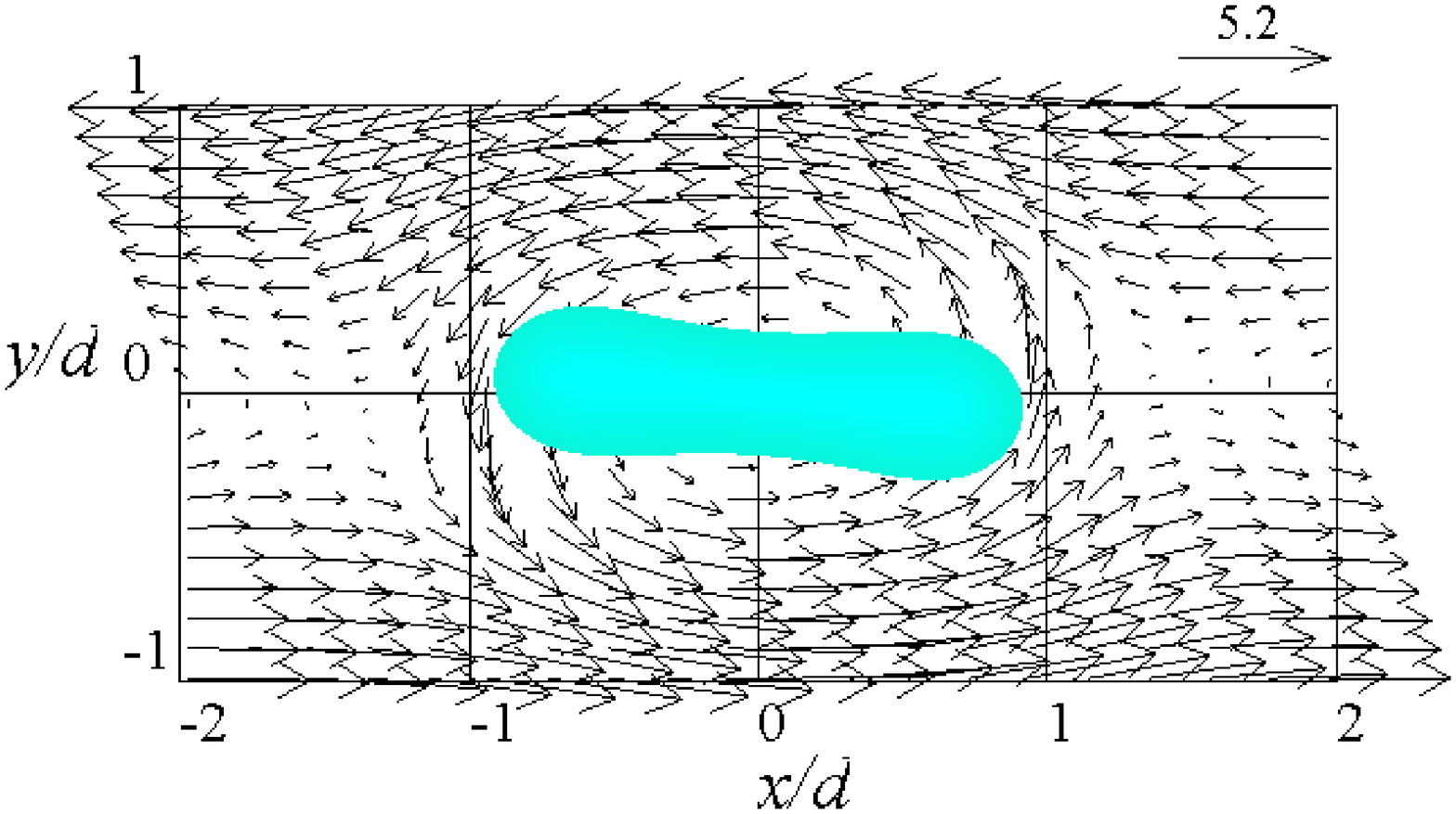} \\
(c) $T=0.3$
\end{minipage}
\hspace{0.02\linewidth}
\begin{minipage}{0.48\linewidth}
\includegraphics[trim=0mm 0mm 0mm 0mm, clip, width=70mm]{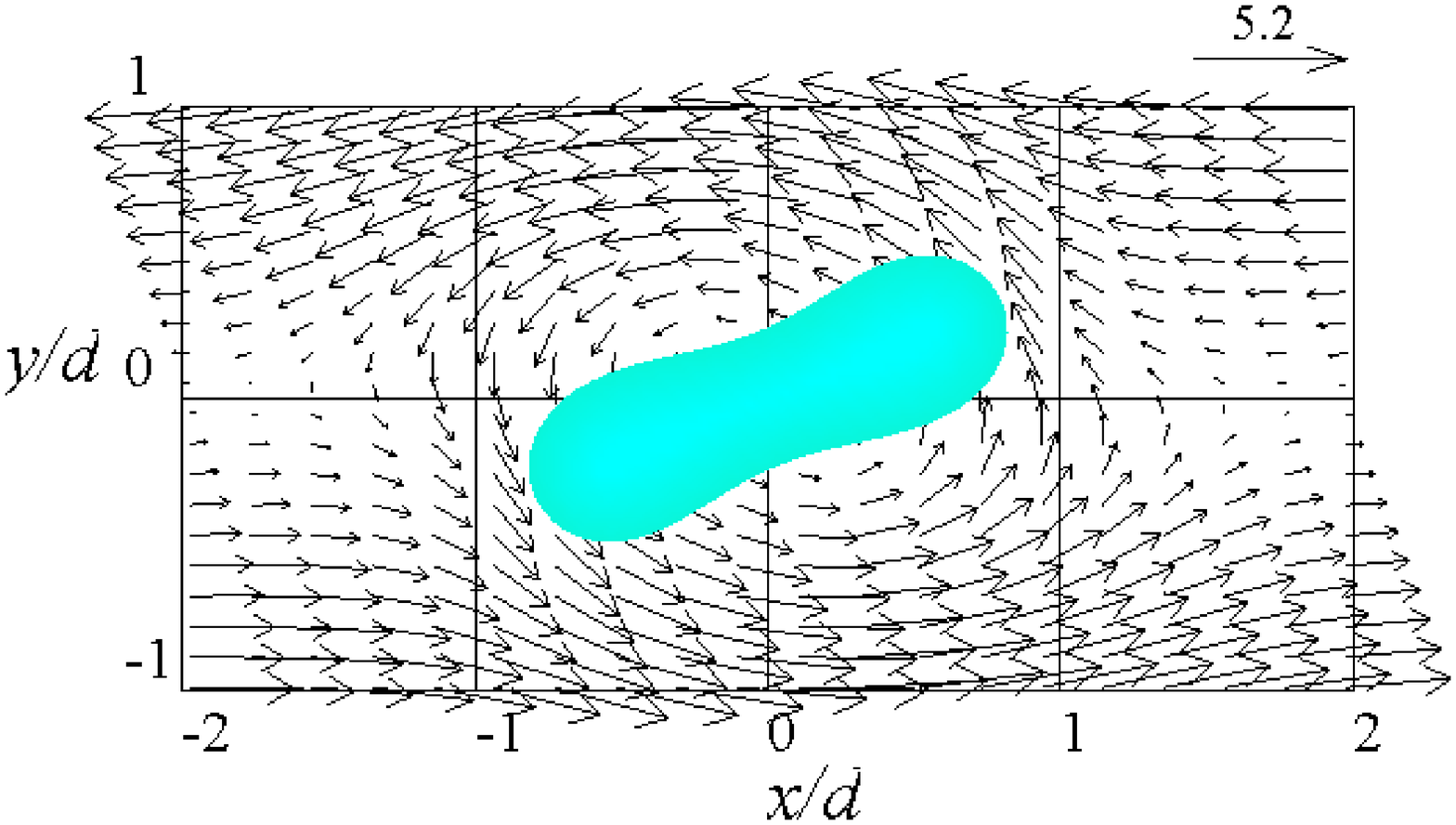} \\
(d) $T=0.5$
\end{minipage}

\begin{minipage}{0.48\linewidth}
\includegraphics[trim=0mm 0mm 0mm 0mm, clip, width=70mm]{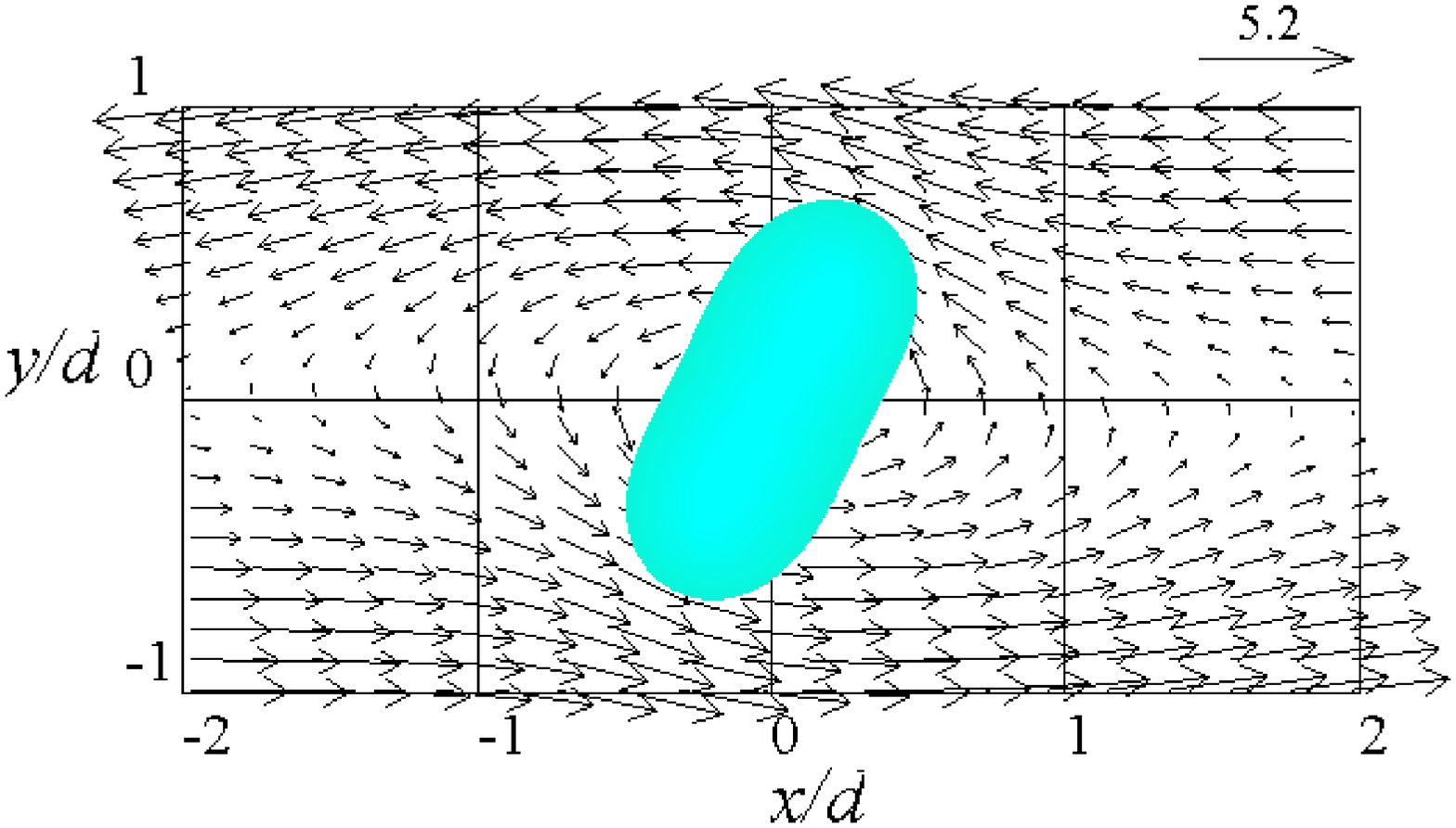} \\
(e) $T=0.7$
\end{minipage}
\hspace{0.02\linewidth}
\begin{minipage}{0.48\linewidth}
\includegraphics[trim=0mm 0mm 0mm 0mm, clip, width=70mm]{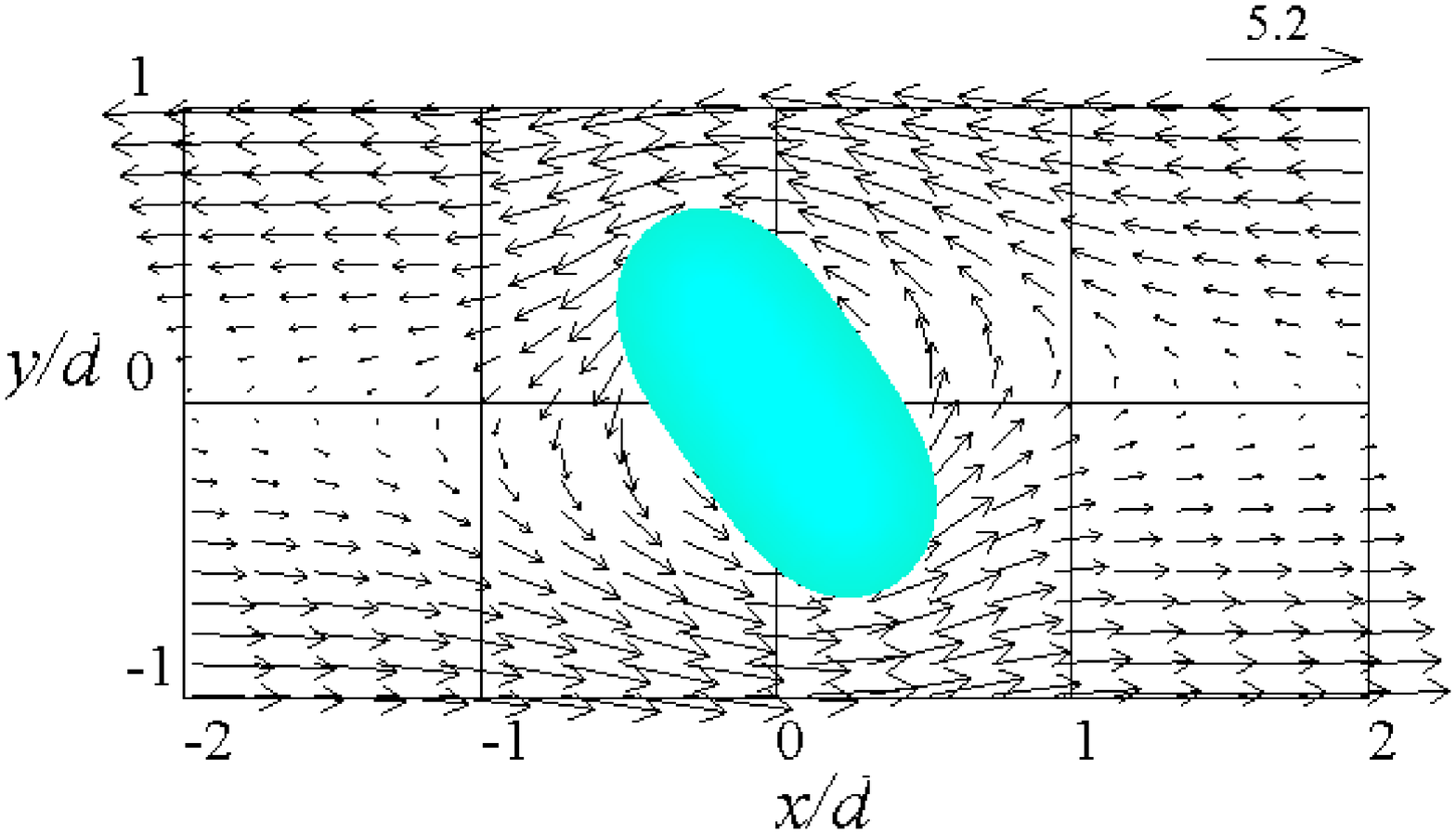} \\
(f) $T = 0.9$
\end{minipage}
\caption{Time variations of ligament interface and velocity vectors 
for $\Delta U^{*}=11.9$}
\label{deltu20}
\end{figure}
%------------------------------------------------------------------------------

%------------------------------------------------------------------------------
% Figure 5
%------------------------------------------------------------------------------
\begin{figure}[!t]
\begin{minipage}{0.48\linewidth}
\includegraphics[trim=0mm 0mm 17mm 0mm, clip, width=62mm]{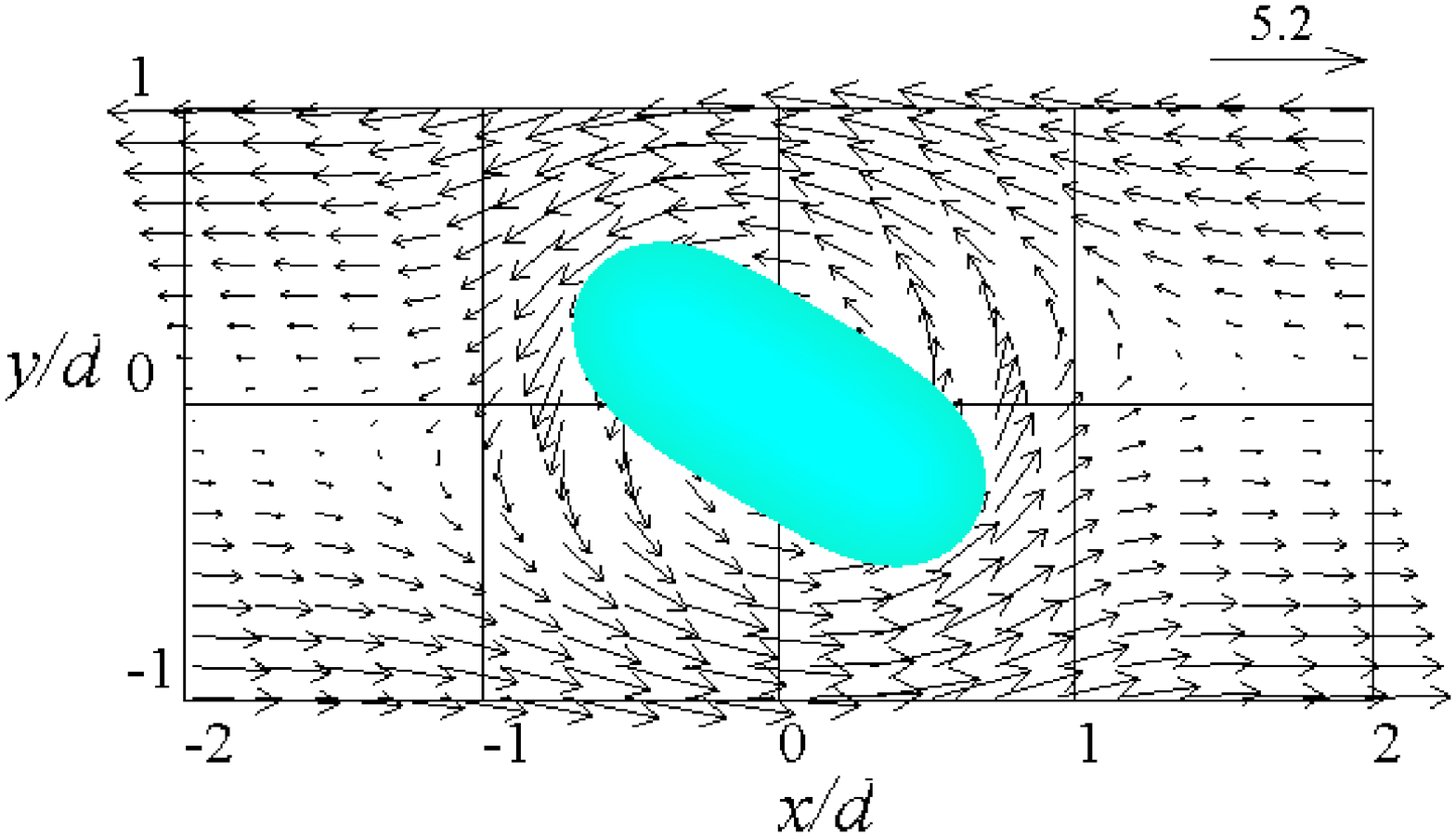} 
\end{minipage}
\hspace{0.02\linewidth}
\begin{minipage}{0.48\linewidth}
\includegraphics[trim=0mm 0mm 0mm -10mm, clip, width=74mm]{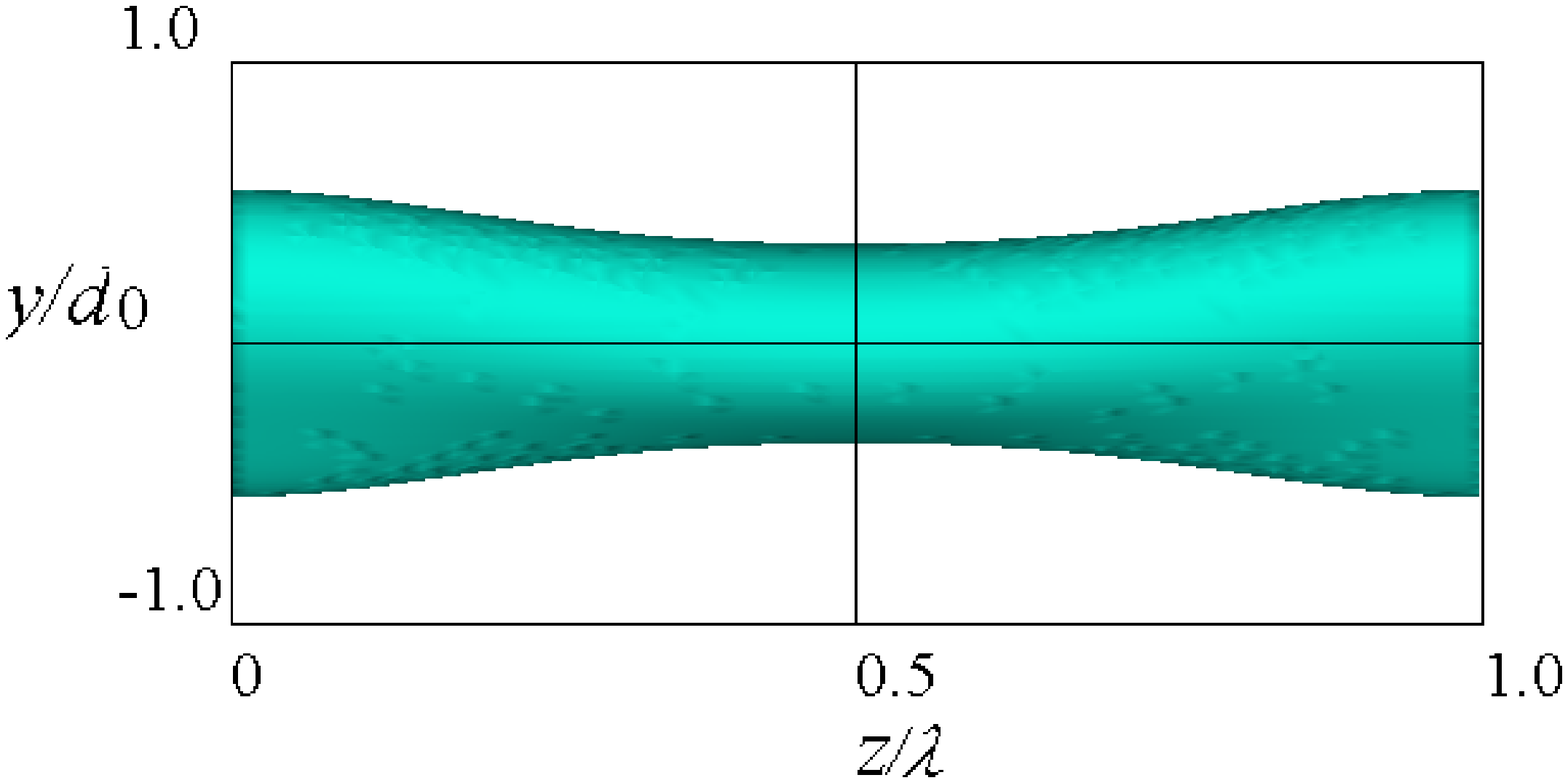}
\end{minipage}
(a) $T=1.0$

\begin{minipage}{0.48\linewidth}
\includegraphics[trim=0mm 0mm 17mm 0mm, clip, width=62mm]{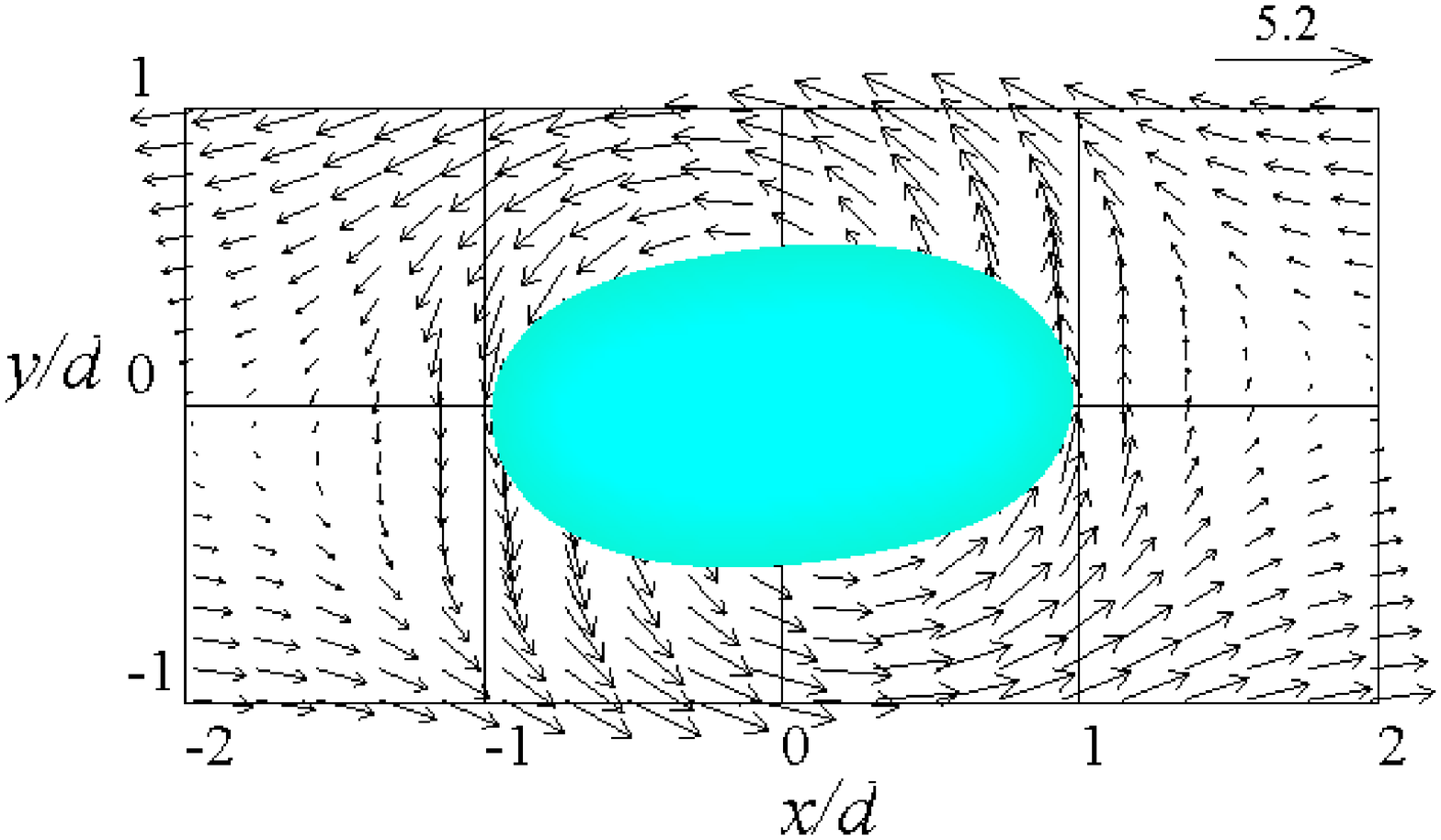} 
\end{minipage}
\hspace{0.02\linewidth}
\begin{minipage}{0.48\linewidth}
\includegraphics[trim=0mm 0mm 0mm -10mm, clip, width=74mm]{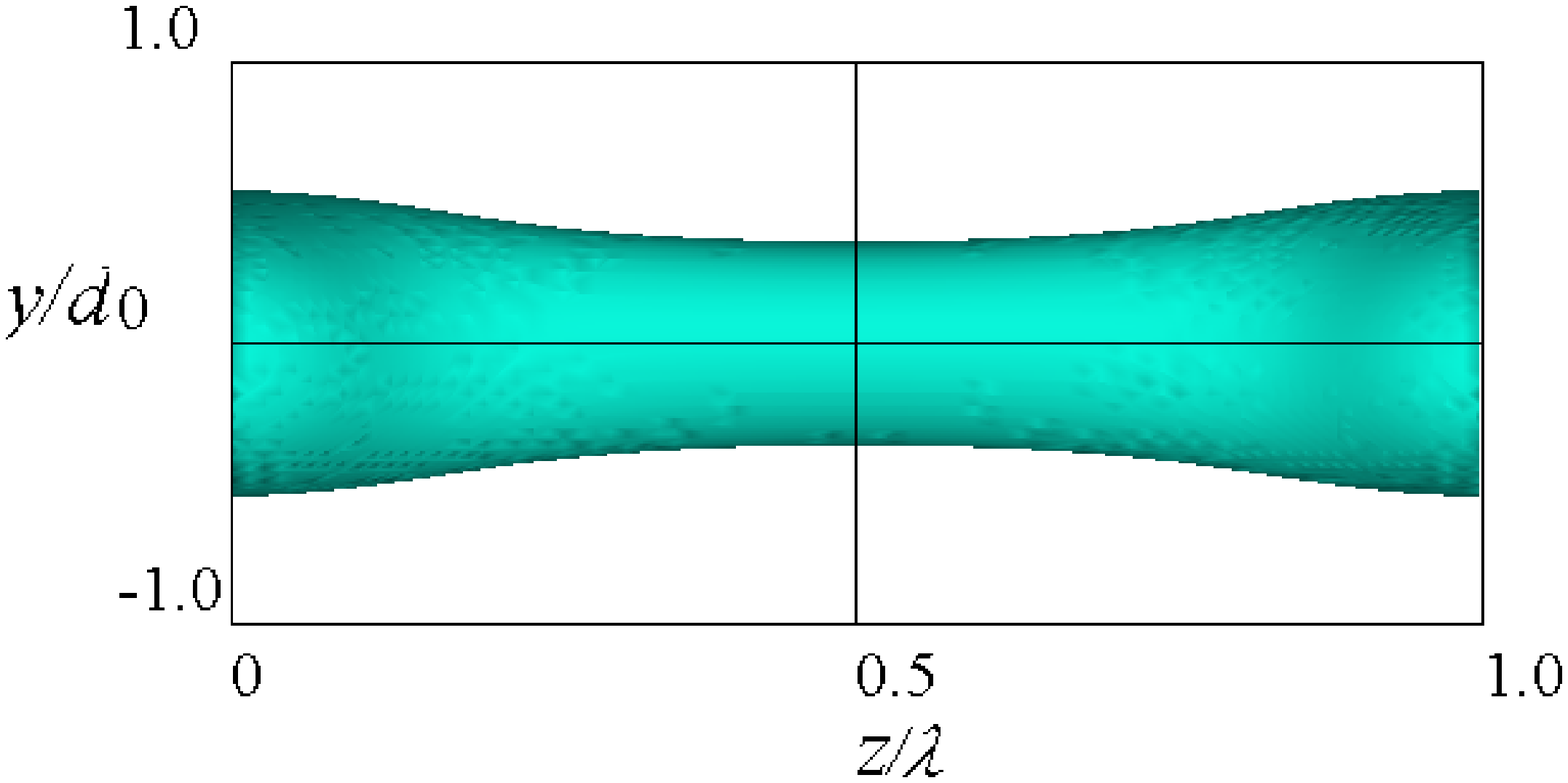}
\end{minipage}
(b) $T=2.0$

\begin{minipage}{0.48\linewidth}
\includegraphics[trim=0mm 0mm 17mm 0mm, clip, width=62mm]{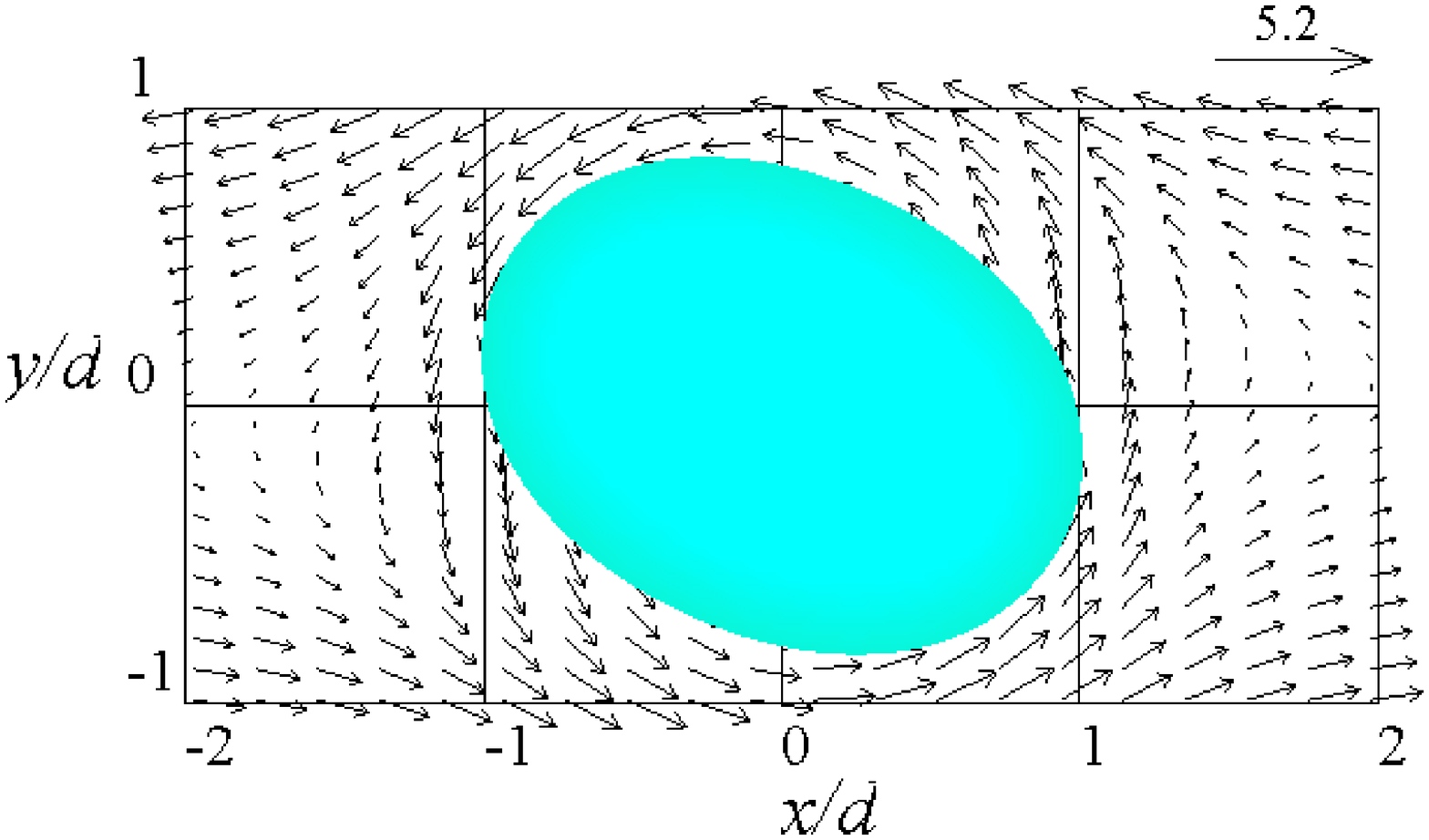} 
\end{minipage}
\hspace{0.02\linewidth}
\begin{minipage}{0.48\linewidth}
\includegraphics[trim=0mm 0mm 0mm -10mm, clip, width=74mm]{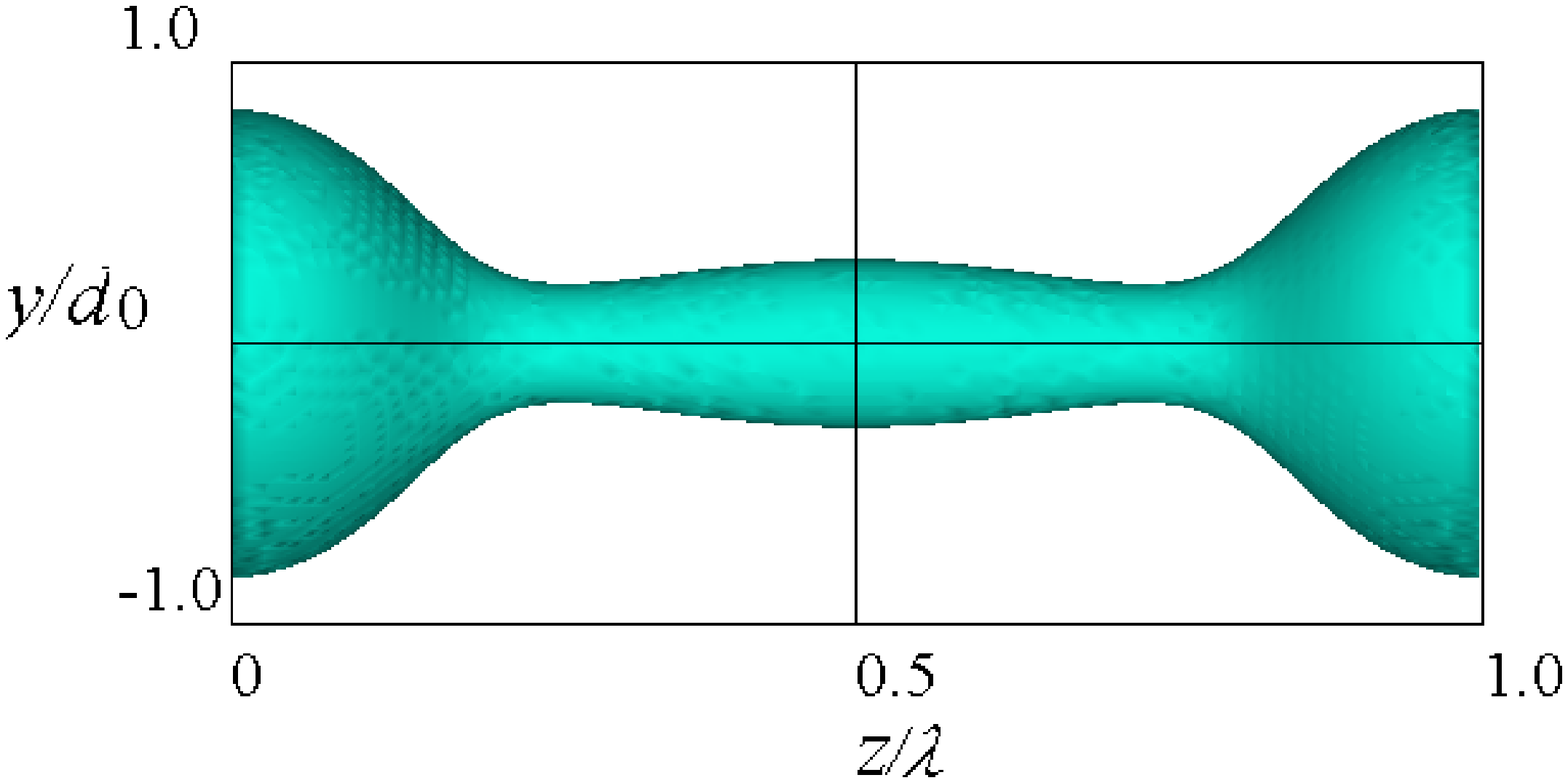}
\end{minipage}
(c) $T=3.0$

\begin{minipage}{0.48\linewidth}
\includegraphics[trim=0mm 0mm 17mm 0mm, clip, width=62mm]{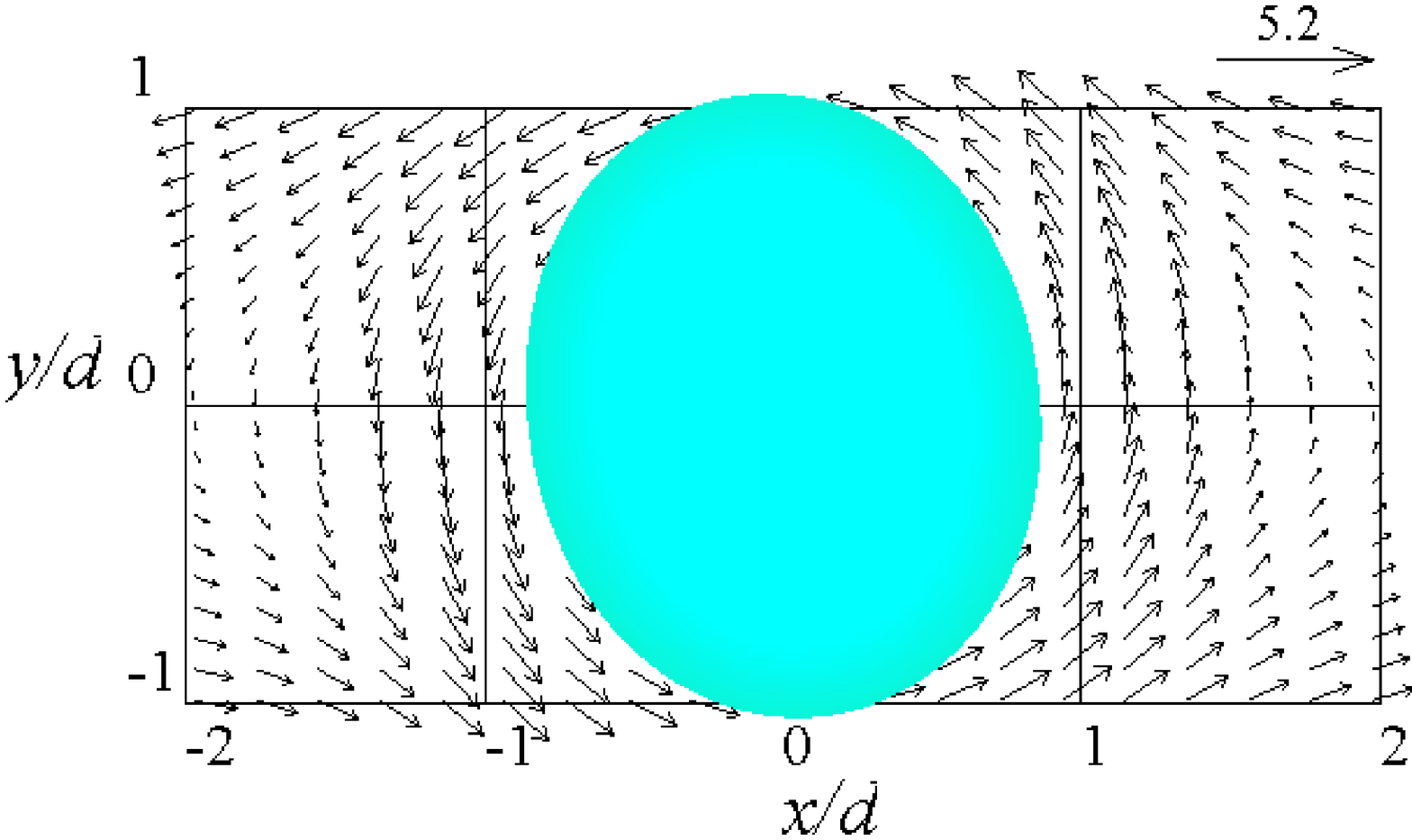} 
\end{minipage}
\hspace{0.02\linewidth}
\begin{minipage}{0.48\linewidth}
\includegraphics[trim=0mm 0mm 0mm -10mm, clip, width=74mm]{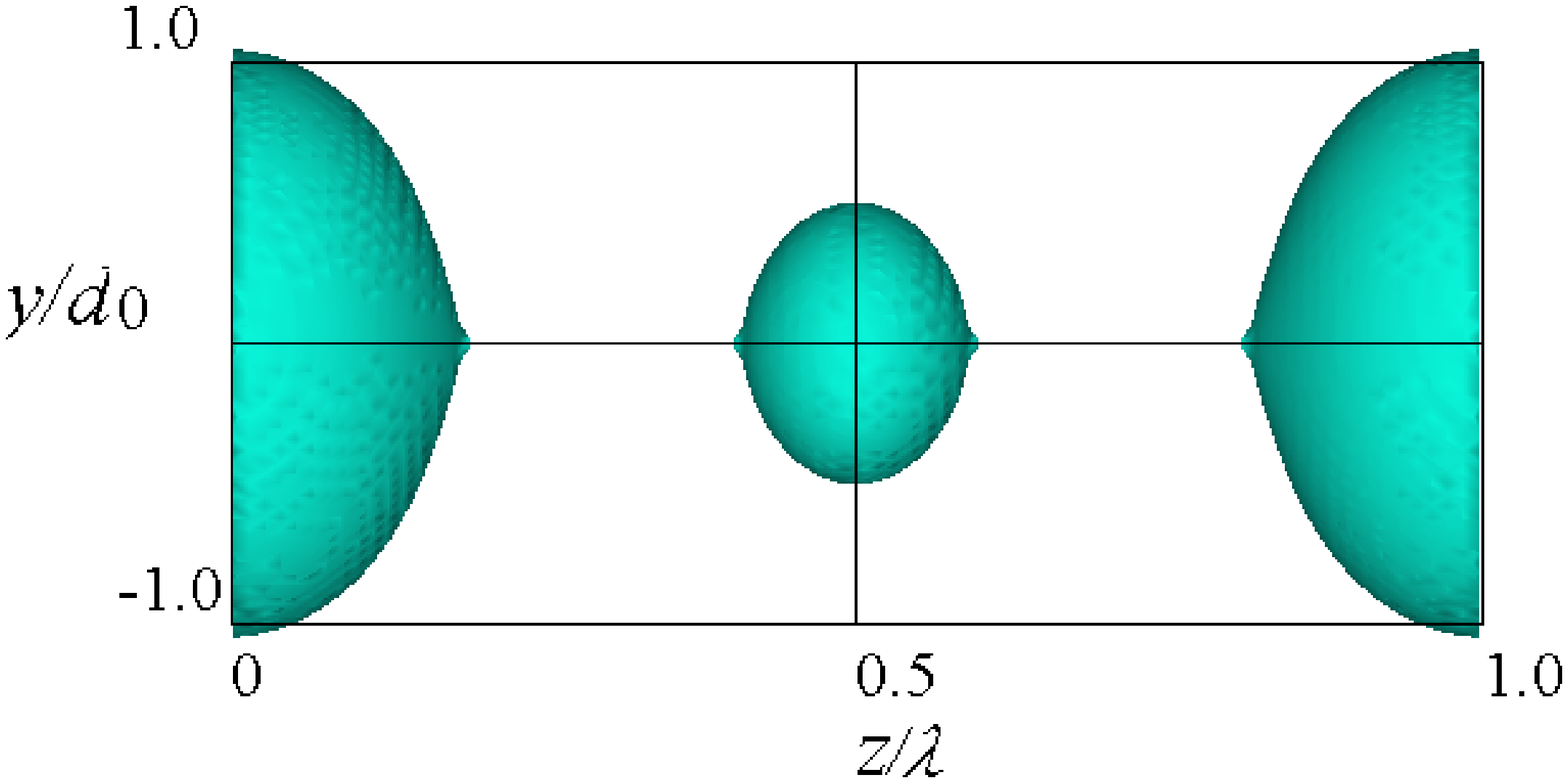}
\end{minipage}
(d) $T=4.0$
\vspace*{-0.5\baselineskip}
\caption{Time variations of ligament interface and velocity vectors 
for $\Delta U^{*}=11.9$}
\label{deltu20_yz}
\end{figure}
%------------------------------------------------------------------------------

%------------------------------------------------------------------------------
% Figure 6
%------------------------------------------------------------------------------
\begin{figure}[!t]
\begin{minipage}{0.325\linewidth}
\includegraphics[trim=2mm 10mm 8mm 6mm, clip, width=45mm]{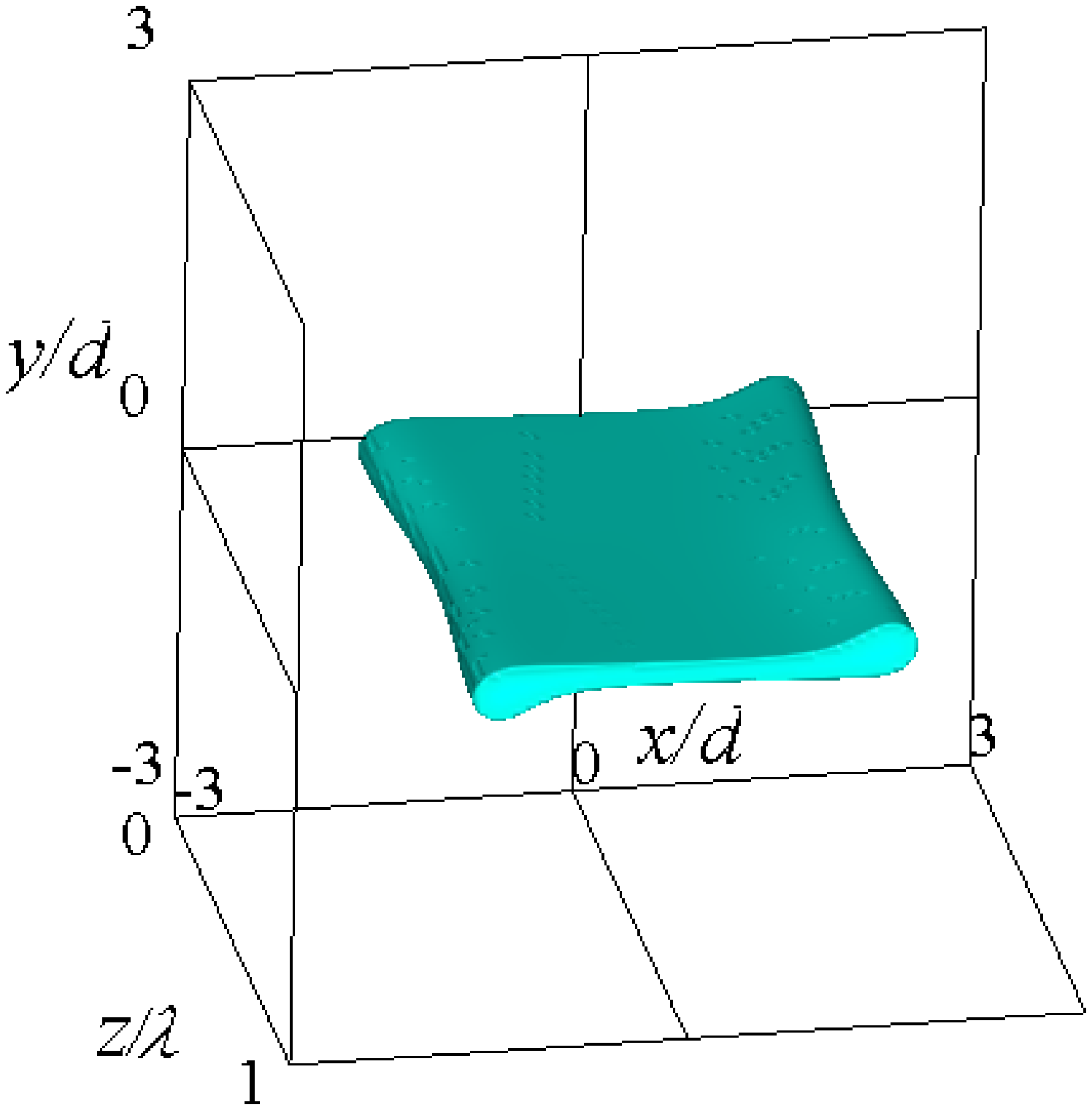} \\
(a) $T=0.3$
\end{minipage}
\begin{minipage}{0.325\linewidth}
\includegraphics[trim=2mm 10mm 8mm 6mm, clip, width=45mm]{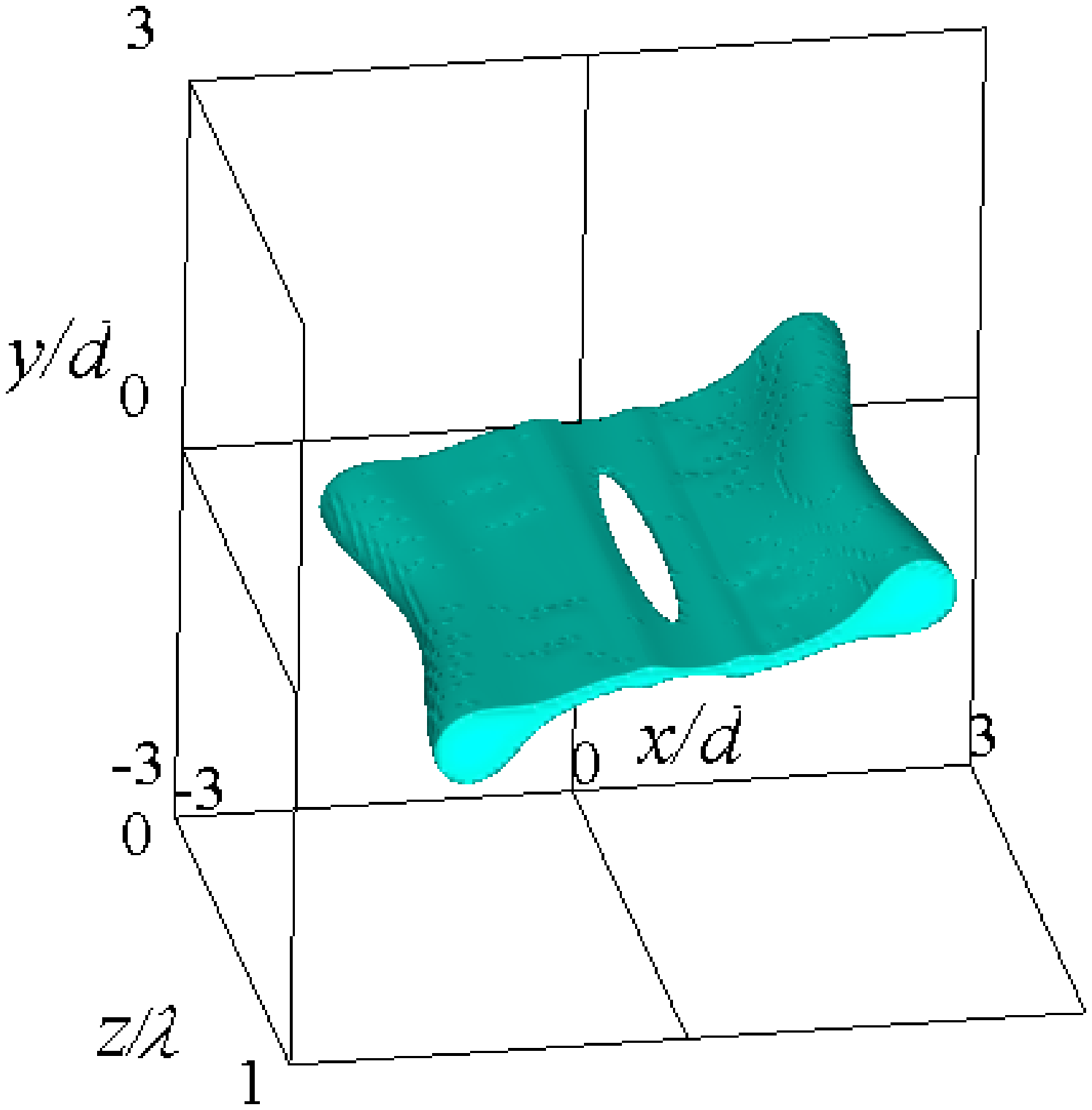} \\
(b) $T=0.6$
\end{minipage}
\begin{minipage}{0.325\linewidth}
\includegraphics[trim=2mm 10mm 8mm 6mm, clip, width=45mm]{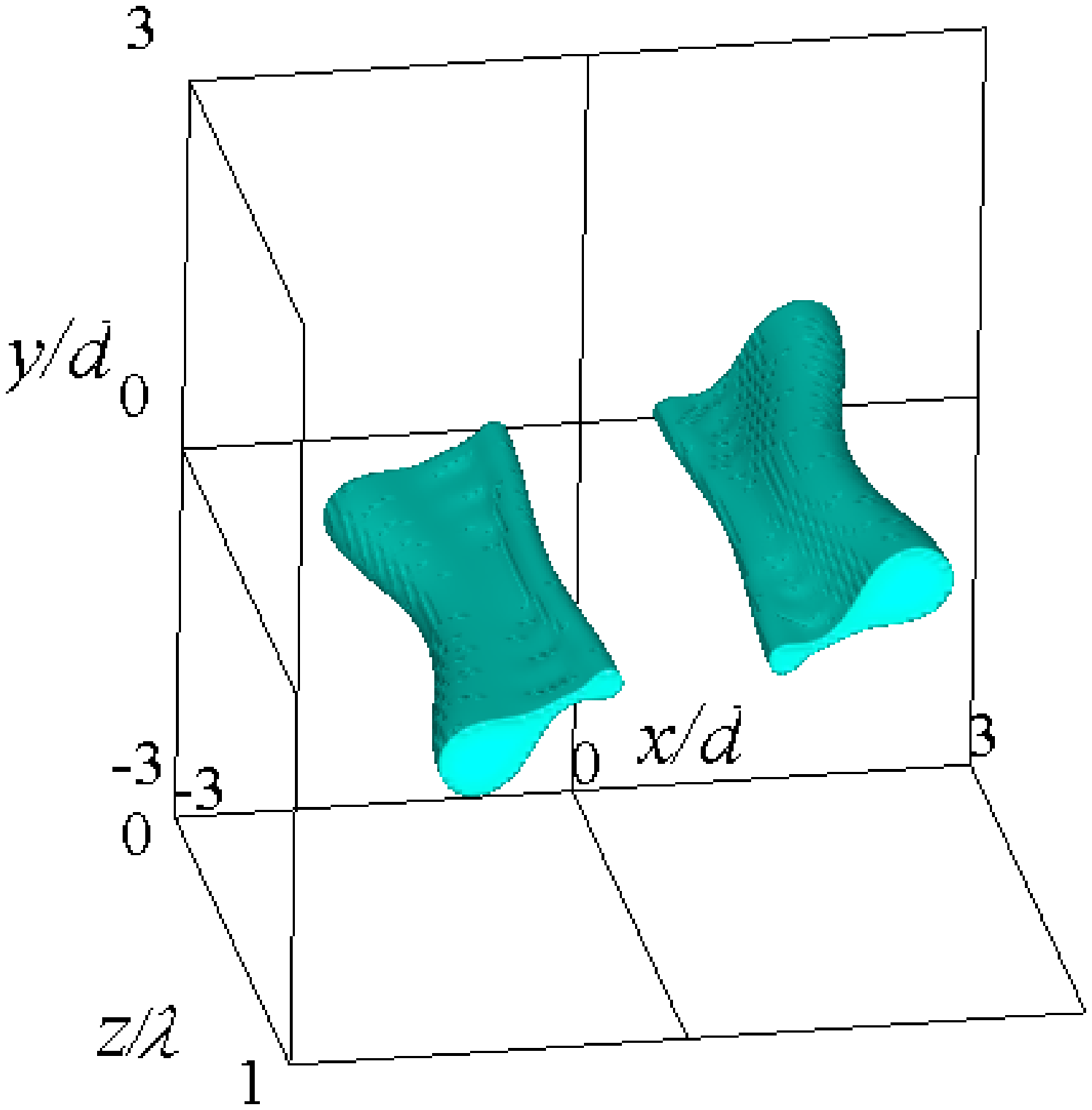} \\
(c) $T=0.7$
\end{minipage}
\begin{minipage}{0.325\linewidth}
\includegraphics[trim=2mm 10mm 8mm 6mm, clip, width=45mm]{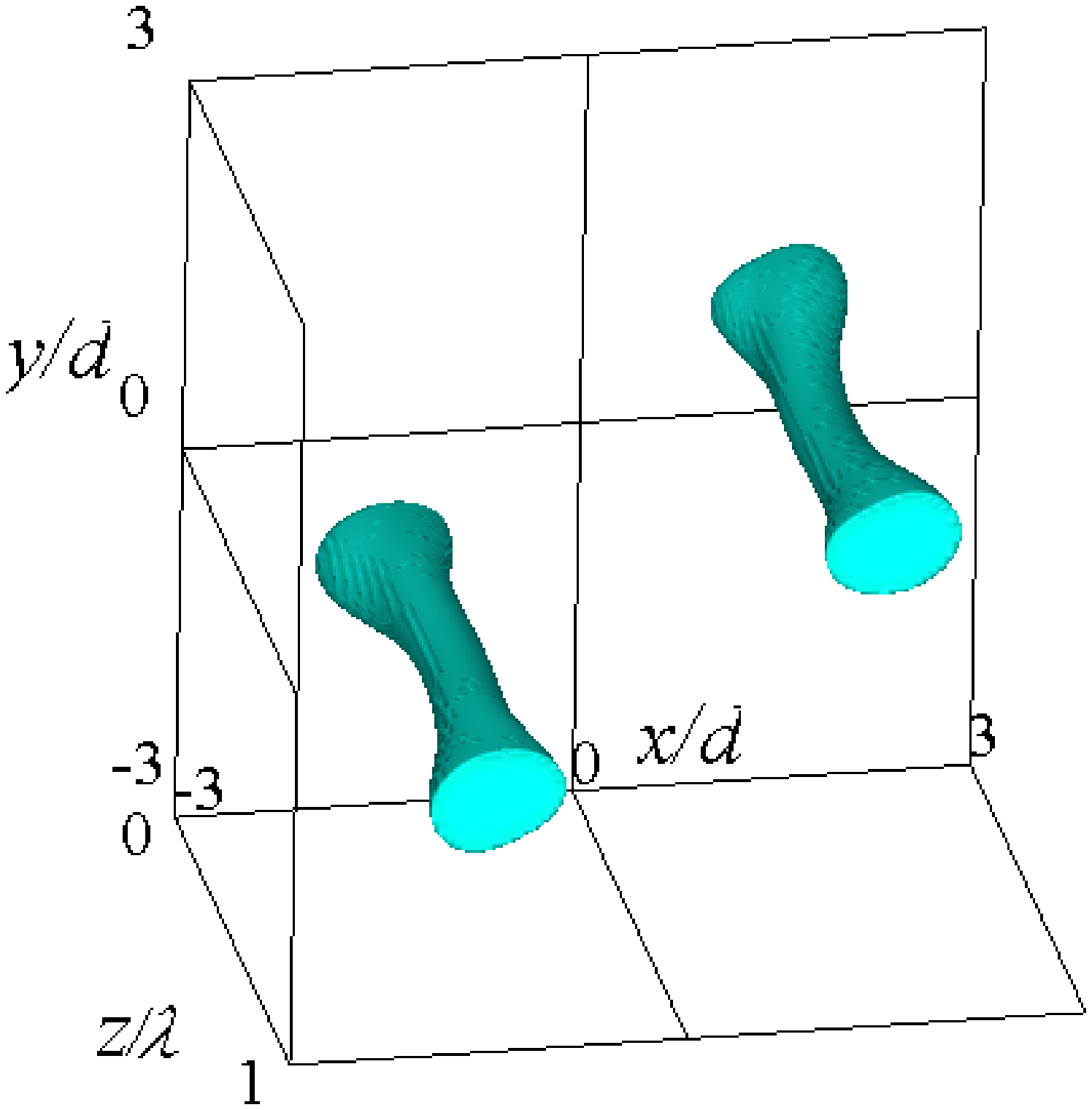} \\
(d) $T=1.2$
\end{minipage}
\begin{minipage}{0.325\linewidth}
\includegraphics[trim=2mm 10mm 8mm 6mm, clip, width=45mm]{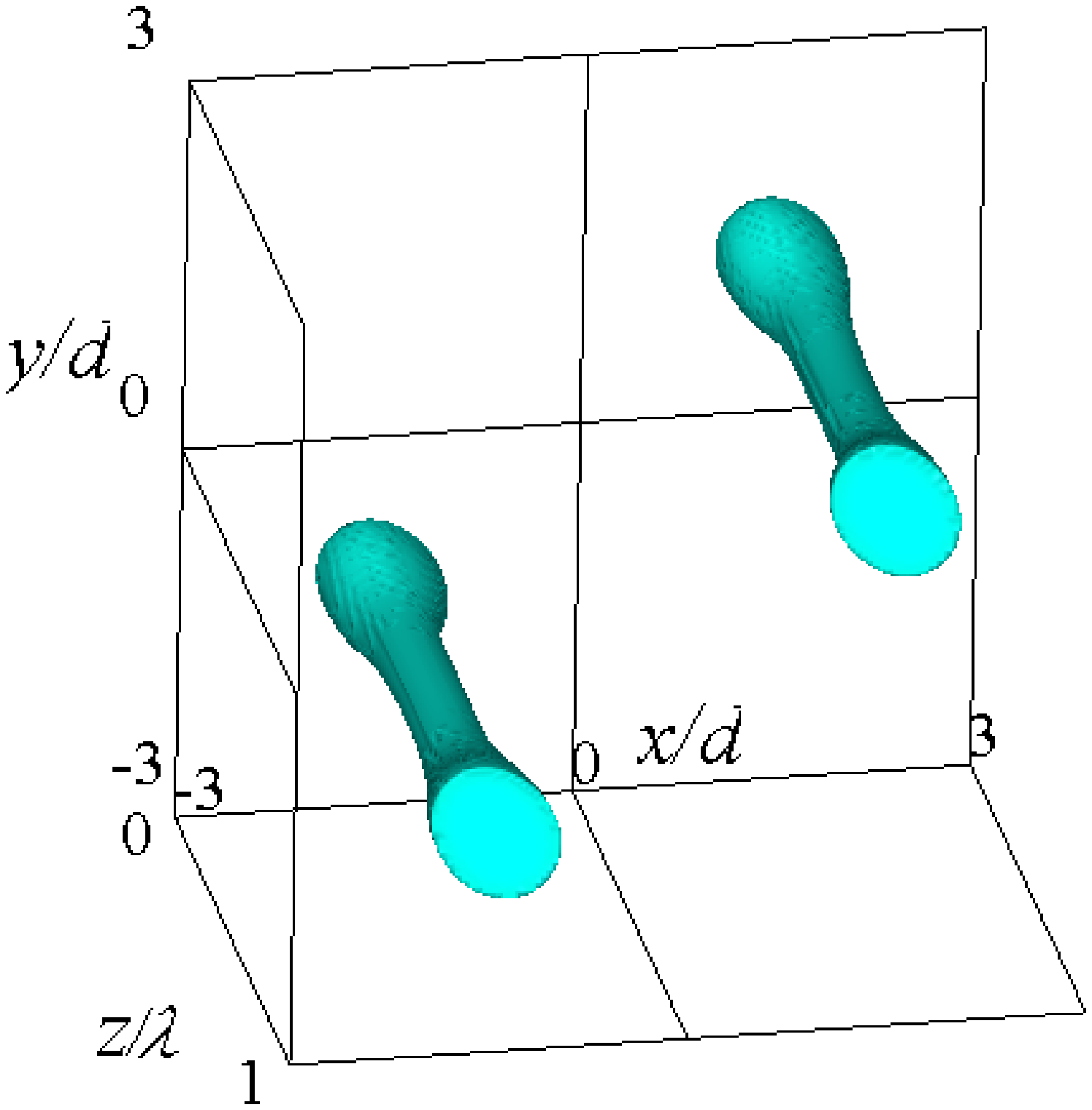} \\
(e) $T=1.5$
\end{minipage}
\begin{minipage}{0.325\linewidth}
\includegraphics[trim=2mm 10mm 8mm 6mm, clip, width=45mm]{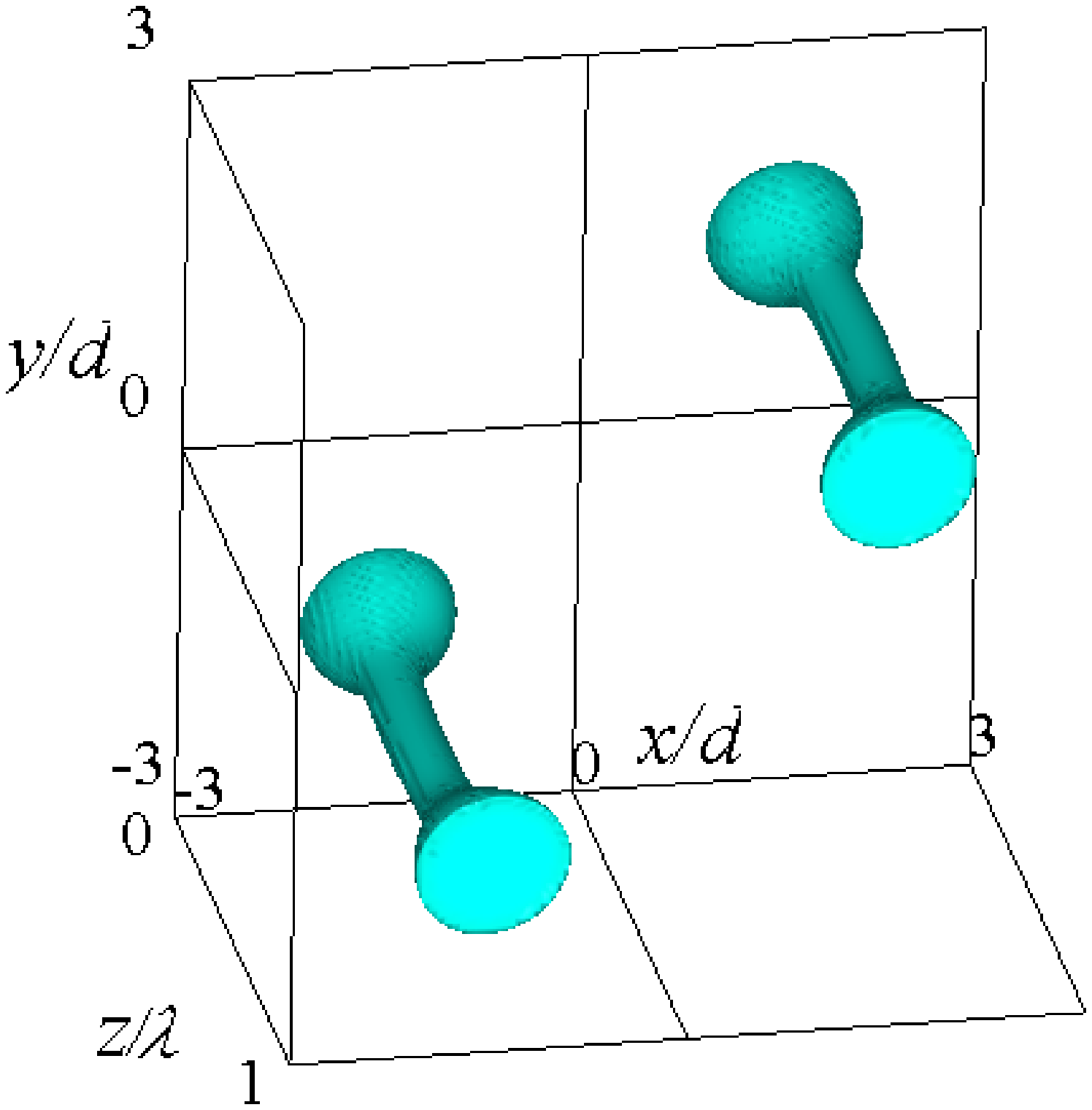} \\
(f) $T=1.8$
\end{minipage}
\caption{Time variations of interface for $\Delta U^{*}=23.7$}
\label{deltu40}
\end{figure}
%------------------------------------------------------------------------------
First, we investigate the deformation of a ligament due to shear flow. 
Figure \ref{deltu20} shows the time variation of the flow field and ligament 
at $ka=0.7$ and $\Delta U^*=11.9$. 
Here, the interface distribution and velocity field near $z/\lambda=1.0$ 
are shown. 
At $T=0-0.3$, the ligament is stretched in the $x$-direction 
by the shear of the airflow. 
At that time, the airflow is along the interface of the ligament. 
As a result, after $T=0.5$, the ligament rotates counterclockwise. 
At $T=0.9$, the cross-section of the ligament is elliptical.

The time variation of the flow field 
and ligament after $T=1.0$ for $ka=0.7$ and $\Delta U^*=11.9$ is shown in Fig. \ref{deltu20_yz}. 
The figure on the left shows the interface distribution and velocity field 
near $z/\lambda=1.0$, 
and the figure on the right shows a side view of the ligament. 
The ligament interface grows with time and the ligament eventually breaks up. 
The position of the ligament interface at $x/d=0$ oscillates up and down. 
The radius at the center of the ligament decreases 
with time. 
In each side view at $T=1.0$ and 2.0, the interface distribution has 
a cosine-like form 
and a constriction is generated at the center of the ligament. 
At $T=3.0$, the interface disturbance grows in a non-cosine manner, 
and the interface near $z/\lambda=0.25$ and 0.75 is constricted. 
As a result, the ligament does not break up in the center; instead, 
it breaks up around $z/\lambda=0.25$ and 0.75. 
At $T=4.0$, a small-scale droplet forms between the droplets 
at both ends. 
The small droplet in the center is called a satellite droplet. 
The formation of satellite droplets has been confirmed 
in previous experiments and numerical analyses for slow jets 
\cite{Goedde&Yuen_1970, Rutland&Jameson_1971, Lakdawala_et_al_2015}. 
In this study, the droplets at the two ends are referred to as the main droplets.

Figure \ref{deltu40} shows the time variation of the ligament interface 
at $\Delta U^*=23.7$. 
At $T=0.3$, the ligament is stretched in the $x$-direction 
by the shear of the airflow and is deformed into a liquid sheet. 
At $T=0.6$, a perforation forms at the center of the liquid sheet 
due to the initial disturbance at the interface. 
The growth of this perforation causes a breakup at $x/d=0$. 
Because the split liquid sheet contracts due to surface tension, 
two ligaments with diameters smaller than that of the original ligament 
are generated at $T=1.2$. 
After $T=1.2$, the interfaces of these ligaments also grow with time. 
It is considered that these ligaments break up again, 
resulting in the formation of smaller droplets 
and an increase in the number of droplets. 
For $\Delta U^*>23.7$, the transformation to a liquid sheet, 
the breakup of the liquid sheet, 
and the subsequent formation of ligaments are similar to 
those shown in Fig. \ref{deltu40}.

\begin{figure}[!t]
\includegraphics[trim=0mm 0mm 0mm 0mm, clip, width=70mm]{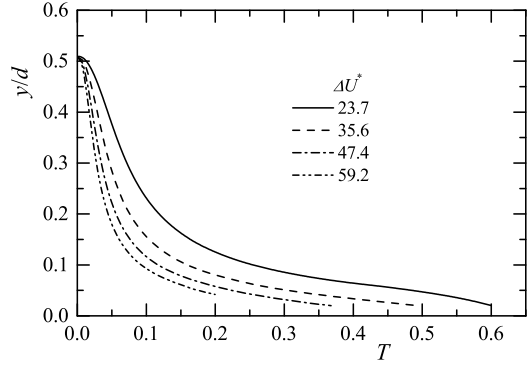}
\vspace*{-0.5\baselineskip}
\caption{Time variations of interface position at $ka=0.7$ 
for $\Delta U^{*}=23.7$, 35.6, 47.4, and 59.2}
\label{int_sgnl_2}
\end{figure}

The time variation of the height of the ligament interface 
at $x/d=0$ near $z/\lambda=1.0$ for $\Delta U^*=23.7$, 35.6, 47.4, and 59.2 is shown in Fig. \ref{int_sgnl_2}. 
Here, the results up to the time when the ligament breakup occurs 
at $x/d=0$ are shown. 
The thickness of the liquid sheet is asymptotic to a constant value 
for all $\Delta U^*$. 
It can be seen that the breakup time for the liquid sheet becomes shorter 
as $\Delta U^*$ increases.

%------------------------------------------------------------------------------
% Figure 7
%------------------------------------------------------------------------------
%------------------------------------------------------------------------------
The ligament formed after the liquid sheet splits becomes unstable again 
under the influence of the shear of the airflow. 
Therefore, in the following, we consider the conditions 
under which the ligament does not split again into liquid sheets 
and investigate the effect of shear flow on the growth rate of the ligament 
and the breakup for $\Delta U^*=0$, 5.9, and 11.9.

%++++++++++++++++++++++++++++++++++++++++++++++++++++++++++++++++++++++++++++++
\subsection{Variation in growth rate of ligament due to shear flow}
%++++++++++++++++++++++++++++++++++++++++++++++++++++++++++++++++++++++++++++++

%------------------------------------------------------------------------------
% Figure 8
%------------------------------------------------------------------------------
\begin{figure}[!t]
\begin{minipage}{0.48\linewidth}
\includegraphics[trim=2mm 0mm 0mm 0mm, clip, width=70mm]{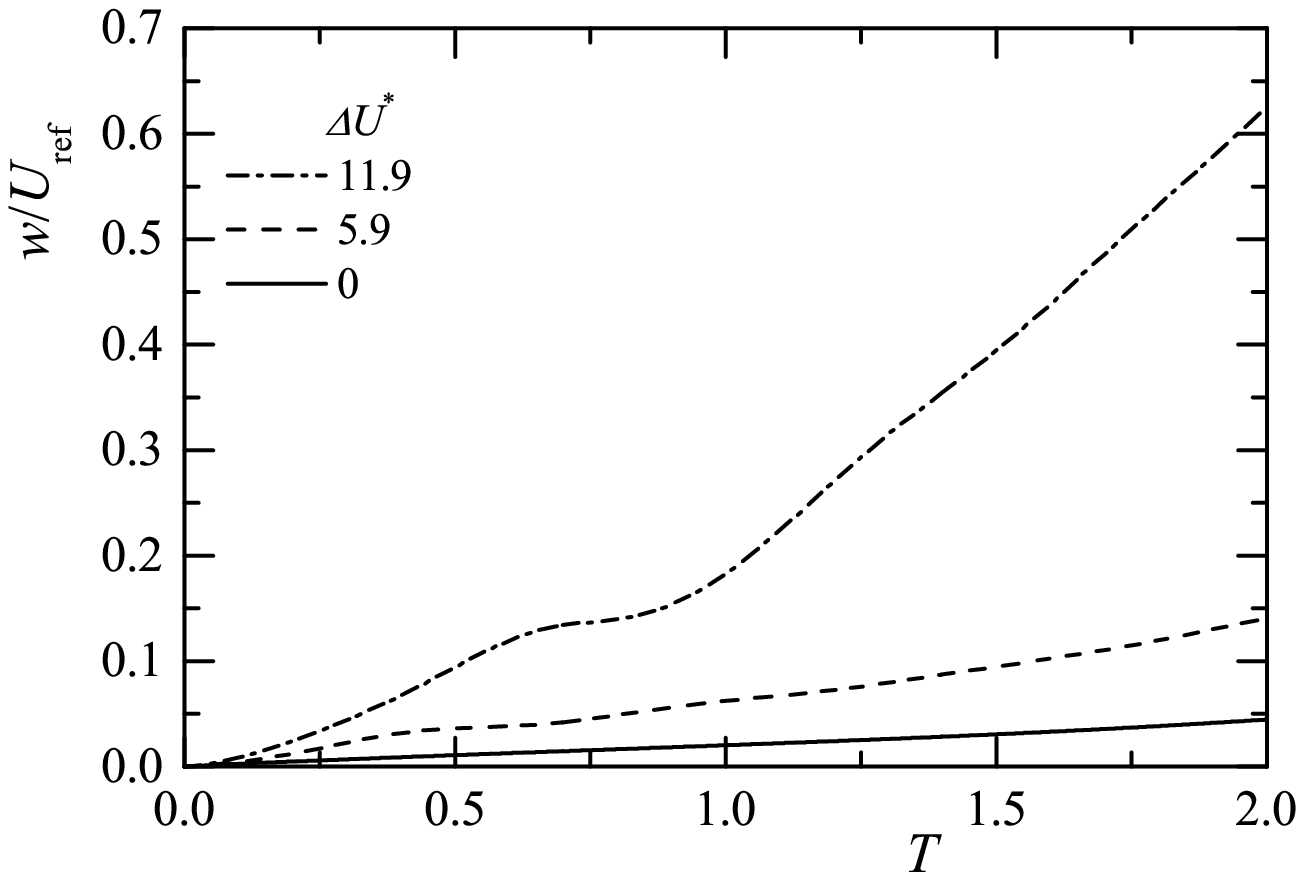} \\
(a) denormalized distributions
\end{minipage}
\hspace{0.02\linewidth}
\begin{minipage}{0.48\linewidth}
\includegraphics[trim=2mm 0mm 0mm 0mm, clip, width=70mm]{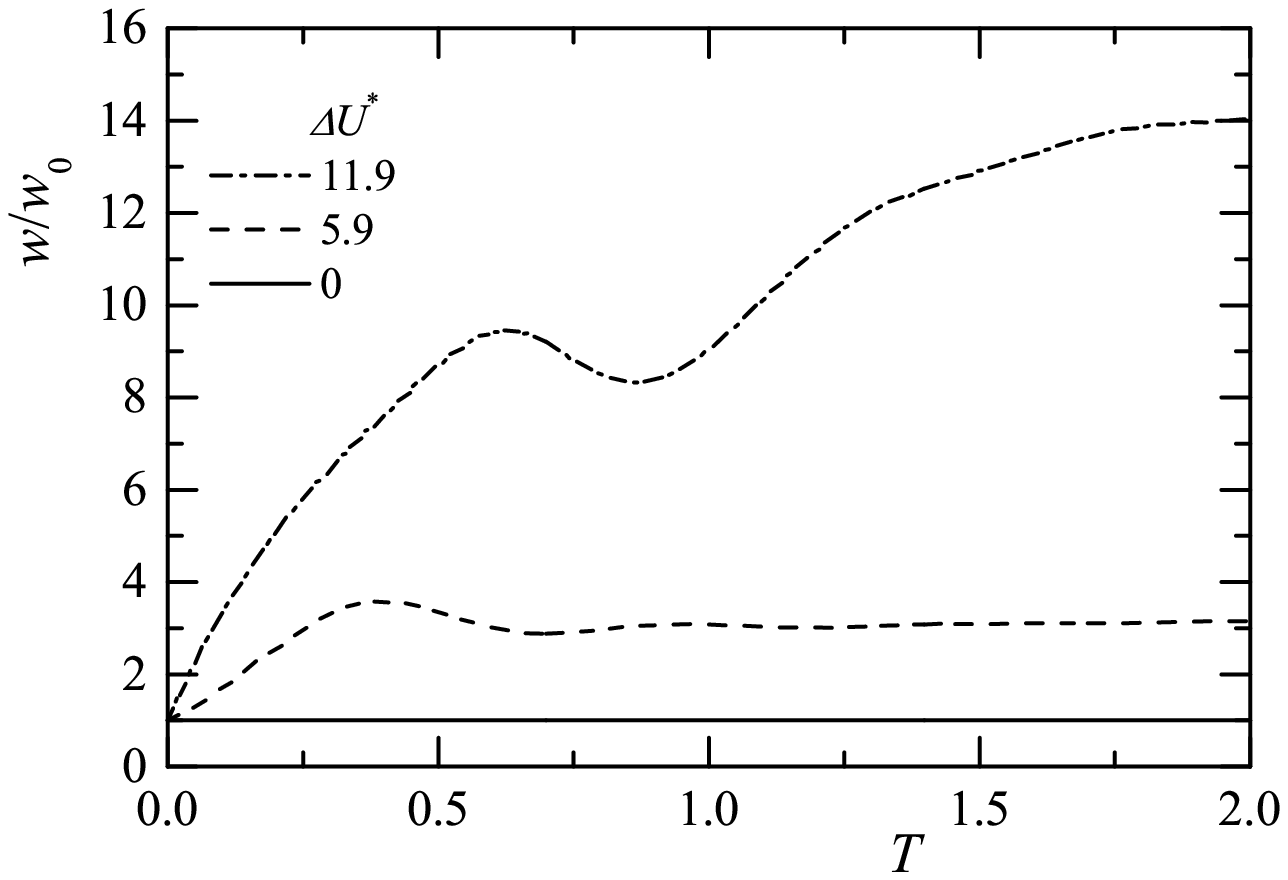} \\
(b) normalized distributions
\end{minipage}
\caption{Time variations of cross--sectional averaged axial velocity 
for $\Delta U^{*}=0$, 5.9, and 11.9}
\label{mw_sgnl}
\end{figure}
%------------------------------------------------------------------------------

To investigate the flow in the ligament, 
the time variation of the axial velocity averaged in the cross-section 
in the ligament at $z/\lambda=0.75$ for $ka=0.7$ and $\Delta U^*=0$, 5.9, and 11.9 is shown in Fig. \ref{mw_sgnl}(a). 
Figure \ref{mw_sgnl}(b) shows the values of the axial velocity $w$ 
in Fig. \ref{mw_sgnl}(a) non-dimensionalized by the axial velocity $w_0$ 
at $\Delta U^*=0$. 
As the interface grows, the liquid near the center of the ligament moves to the end 
of the ligament. 
As a result, in Fig. \ref{mw_sgnl}(a), $w$ increases with time. 
In addition, $w$ at each time increases as $\Delta U^*$ increases. 
It can be seen from this result that the velocity of the liquid 
moving from the central part of the ligament to the end part increases, 
and the growth of the interface becomes faster.
In Fig. \ref{mw_sgnl}(b), $w$ at $\Delta U^*=5.9$ and 11.9 is
significantly higher than that at $\Delta U^*=0$. 
The distributions at $\Delta U^*=5.9$ and 11.9 have maximum values 
around $T=0.4$ and 0.6, respectively. 
This is because cross-sections of the ligament at $\Delta U^*=5.9$ and 11.9 
become elliptical between $T=0-0.4$ and $T=0-0.6$, respectively, 
and the destabilization increased. 
At $\Delta U^*=5.9$, the cross-section gradually becomes 
circular again, 
so the increase in $w$ subsides. 
However, at $\Delta U^*=11.9$, the cross-section of the ligament 
is elliptical even after $T=0.6$, so $w$ increases significantly.

%------------------------------------------------------------------------------
% Figure 9
%------------------------------------------------------------------------------
\begin{figure}[!t]
\includegraphics[trim=0mm 0mm 0mm 0mm, clip, width=80mm]{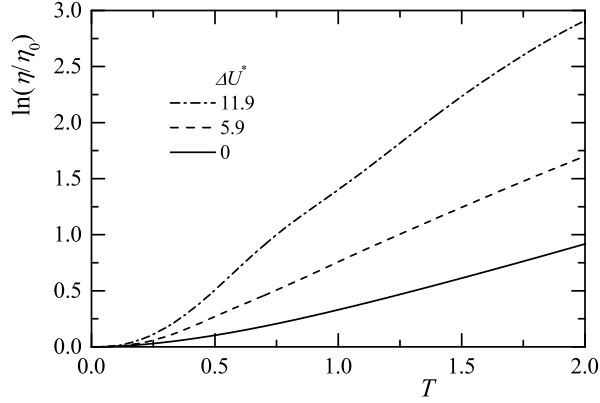}
\vspace*{-0.5\baselineskip}
\caption{Time variations of amplitude at $\Delta U^{*}=0$, 5.9, and 11.9 
for $ka=0.7$ }
\label{amp_3d}
\end{figure}
%------------------------------------------------------------------------------

Figure \ref{amp_3d} shows the time variation of the amplitude of the disturbance 
at the interface for $ka=0.7$ and $\Delta U^*=0$, 5.9, and 11.9 
to investigate the growth rate of the ligament interface. 
Because the cross-section of the ligament affected by the shear flow 
is non-circular, the amplitude $\eta/d$ of the ligament interface 
is calculated using the following equation:
%------------------------------------------------------------------------------
\begin{equation}
   \eta / d = r_\mathrm{eq} / d - r_0 / d,
\end{equation}
%------------------------------------------------------------------------------
where $r_\mathrm{eq}$ is the equivalent radius of the circle 
obtained from the cross-sectional area of the ligament, 
$r_0$ is the radius of the ligament when no initial disturbance is applied, 
and $r_0/d=0.5$. 
The dimensionless growth rates at $\Delta U^*=0$, 5.9, and 11.9 
are 0.22, 0.30, and 0.61, respectively, 
and the growth of the disturbance amplitude becomes faster 
as $\Delta U^*$ increases.

%------------------------------------------------------------------------------
% Figure 10
%------------------------------------------------------------------------------
\begin{figure}[!t]
\includegraphics[trim=0mm 0mm 0mm 0mm, clip, width=80mm]{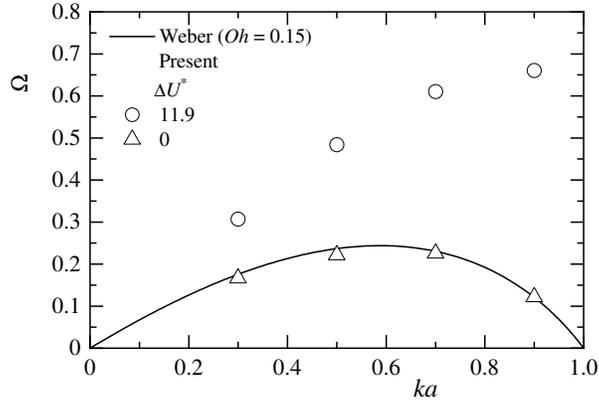}
\vspace*{-0.5\baselineskip}
\caption{Growth rate of interface for $\Delta U^{*}=0$ and 11.9}
\label{grate_3d}
\end{figure}
%------------------------------------------------------------------------------
In Fig. \ref{grate_3d}, the variation of the dimensionless growth rate 
$\Omega$ with the wavenumber $ka$ at $\Delta U^*=0$ and 11.9 
is compared with the theoretical value \cite{Weber_1931}. 
The growth rate of the ligament interface at $\Delta U^*=0$ 
agrees well with the theoretical value 
and it is the highest at $ka=0.7$. 
In contrast, at $\Delta U^*=11.9$, the growth rate is higher than 
the theoretical value, 
and that at $ka=0.9$ is higher than that at $ka=0.7$. 
In a previous study \cite{Rayleigh_1878}, 
the growth rate of a liquid column in a stationary fluid was the highest at $ka=0.7$. 
Thus, the wavenumber that governs the breakup of a liquid column was $ka=0.7$. 
However, in a shear flow, the growth rate of the ligament is higher than 
the theoretical value, 
and increases with increasing wavenumber. 
Therefore, the formed droplet diameter in the shear flow decreases 
due to the increase in the wavenumber that governs the breakup of the ligament, 
and because the growth rate increases, the breakup time for the ligament 
becomes shorter.

%++++++++++++++++++++++++++++++++++++++++++++++++++++++++++++++++++++++++++++++
\subsection{Effect of shear flow on ligament breakup}
%++++++++++++++++++++++++++++++++++++++++++++++++++++++++++++++++++++++++++++++

%------------------------------------------------------------------------------
% Figure 11
%------------------------------------------------------------------------------
\begin{figure}[!t]
\begin{minipage}{0.48\linewidth}
\includegraphics[trim=0mm 7mm 0mm 0mm, clip, width=70mm]{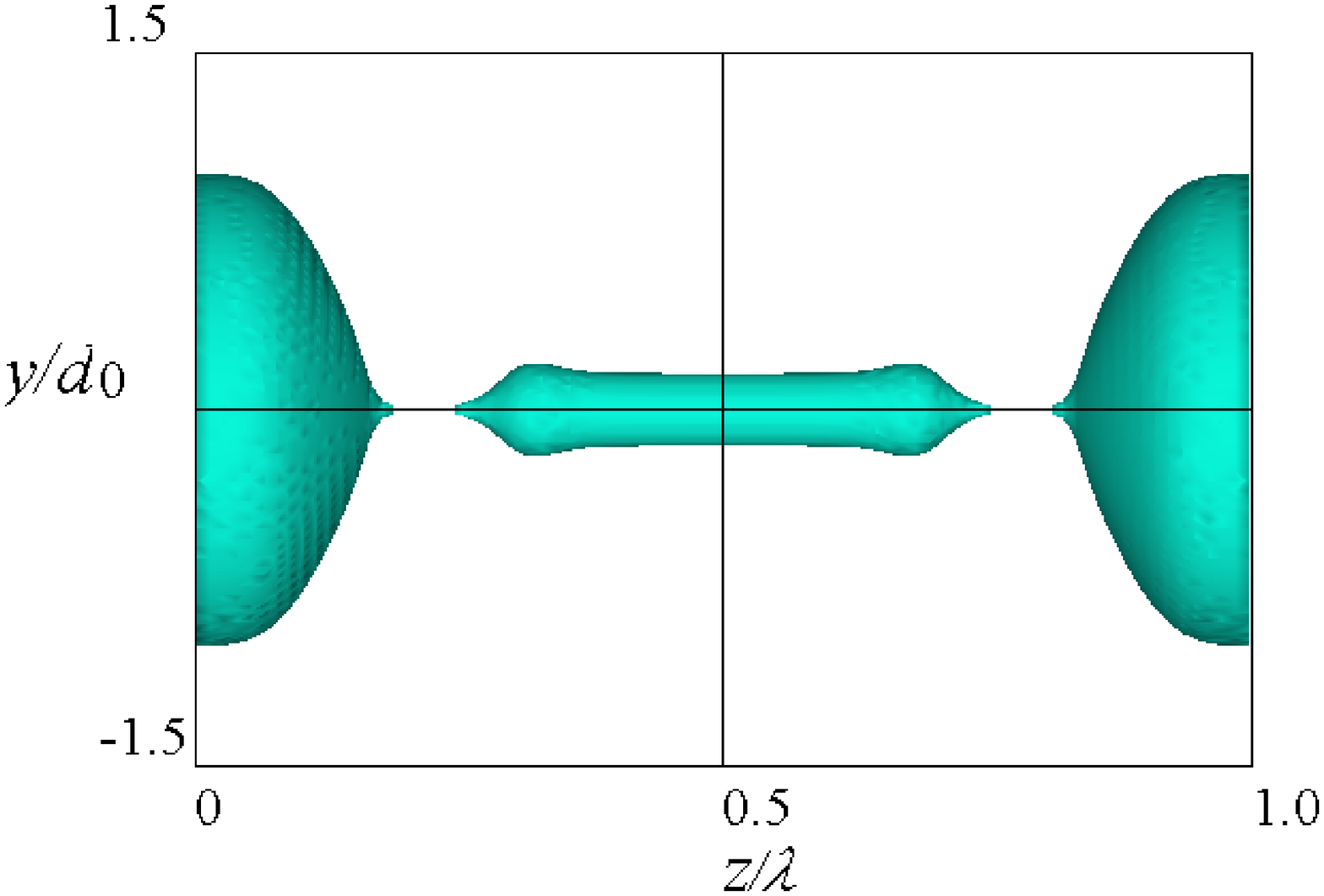} \\
(a) $ka=0.3$ ($T=7.2$)
\end{minipage}
\hspace{0.02\linewidth}
\begin{minipage}{0.48\linewidth}
\includegraphics[trim=0mm 7mm 0mm 0mm, clip, width=70mm]{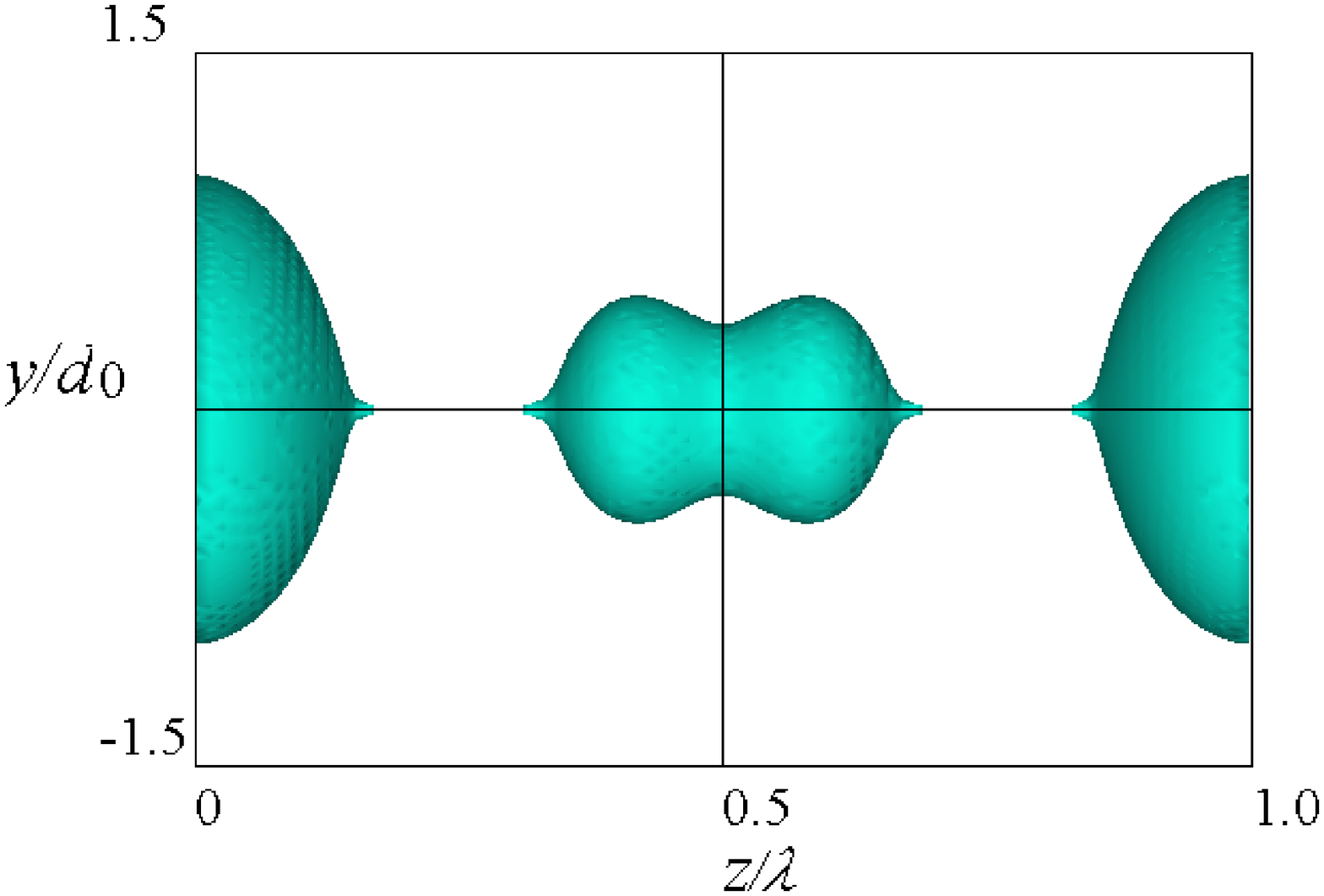} \\
(b) $ka=0.5$ ($T=4.8$)
\end{minipage}
\begin{minipage}{0.48\linewidth}
\includegraphics[trim=0mm 7mm 0mm 0mm, clip, width=70mm]{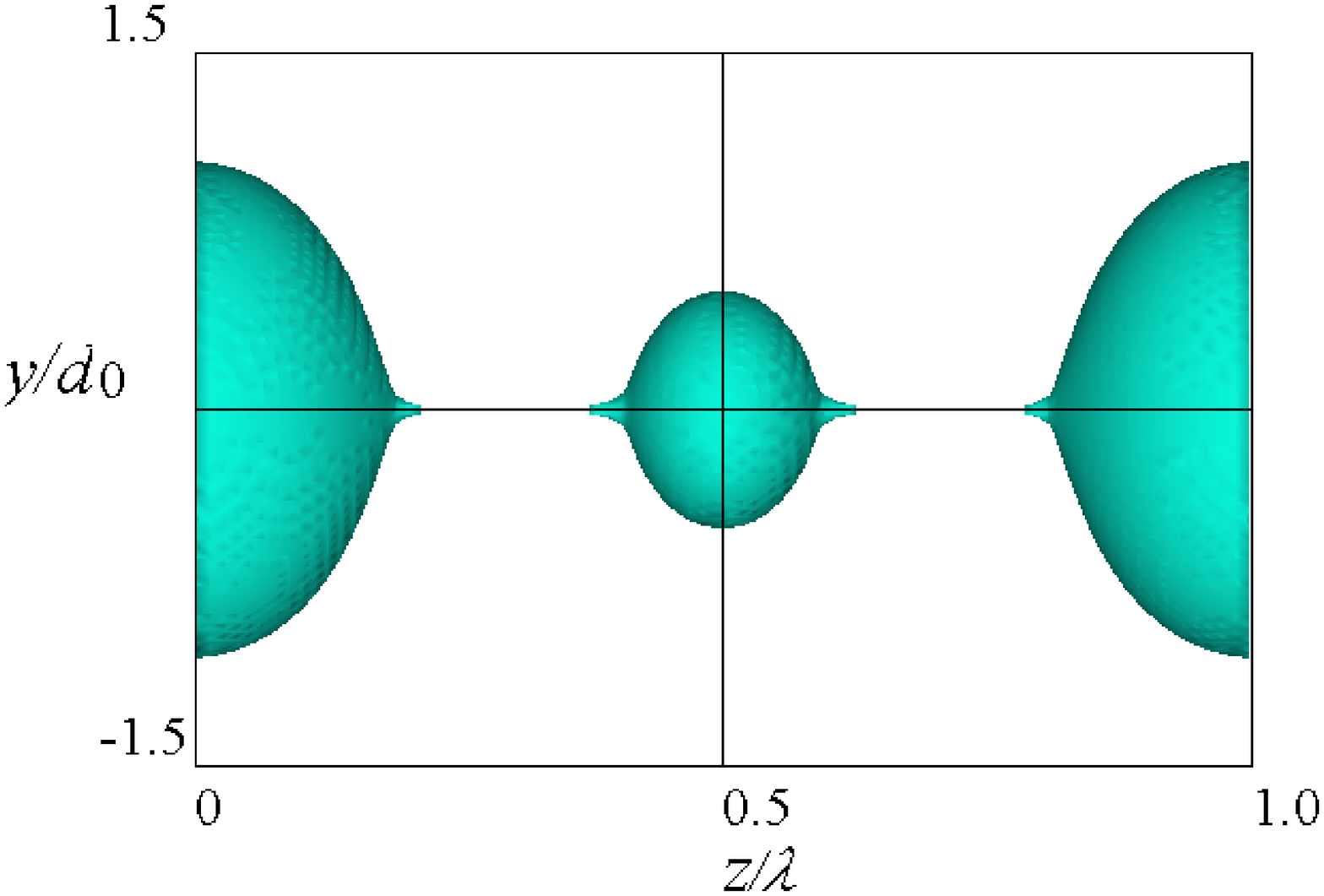} \\
(c) $ka=0.7$ ($T=3.9$)
\end{minipage}
\hspace{0.02\linewidth}
\begin{minipage}{0.48\linewidth}
\includegraphics[trim=0mm 7mm 0mm 0mm, clip, width=70mm]{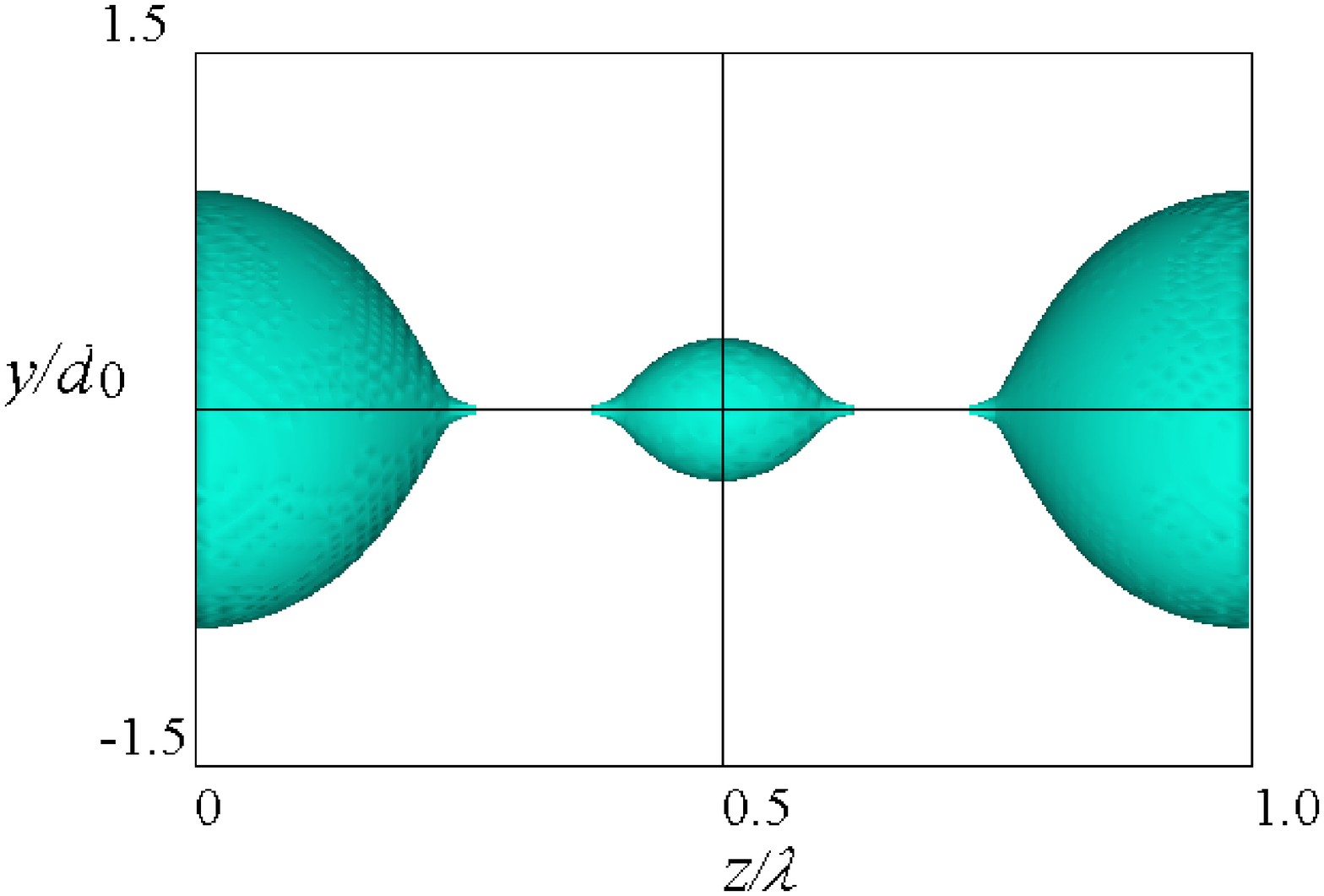} \\
(d) $ka=0.9$ ($T=3.7$)
\end{minipage}
\caption{Side view of ligament interface for $\Delta U^{*}=11.9$}
\label{bup_each_ka}
\end{figure}
%------------------------------------------------------------------------------

Next, we investigate the effect of shear flow on ligament breakup. 
Figure \ref{bup_each_ka} shows the interface distribution immediately 
after breakup at wavenumbers $ka=0.3$, 0.5, 0.7, and 0.9 for $\Delta U^*=11.9$. 
The dimensionless time at each wavenumber is $T=7.2$, 4.8, 3.9, and 3.7, 
respectively. 
Satellite droplets are generated at all wavenumbers. 
At $ka=0.3$, a ligament forms between the main droplets. 
This ligament contracts with time and eventually becomes a droplet. 

%------------------------------------------------------------------------------
% Figure 12
%------------------------------------------------------------------------------
\begin{figure}[!t]
\includegraphics[trim=0mm 0mm 0mm 0mm, clip, width=80mm]{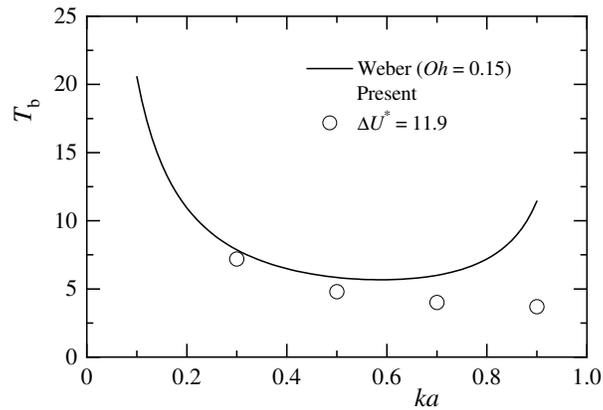}
\vspace*{-0.5\baselineskip}
\caption{Breakup time variations at various wavenumbers 
for $\Delta U^{*}=11.9$}
\label{btime}
\centering
\end{figure}
%------------------------------------------------------------------------------

For $\Delta U^*=11.9$, the time at which breakup was confirmed 
at each wavenumber was taken as the breakup time $T_b$ in this calculation. 
A comparison with the breakup time for the ligament 
predicted using linear theory \cite{Weber_1931} is shown in Fig. \ref{btime}. 
The predicted value was obtained by assuming that the ligament breaks up 
when the interface amplitude at each wavenumber grows 
and becomes equal to the radius of the ligament. 
In this analysis, the growth rate of the ligament at $\Delta U^*=11.9$ 
is higher than the theoretical value \cite{Weber_1931}. 
Therefore, the breakup time for the ligament in this analysis 
is shorter than that predicted by theory.

Table \ref{dsize_eachu} shows the diameter $d_m$ of the main droplet 
and the diameter $d_s$ of the satellite droplet at $\Delta U^*=0$, 
5.9, and 11.9 for $ka=0.7$. 
In this analysis, the equivalent diameter of the droplet was calculated 
from the volume of the liquid in the main droplet and the satellite droplet. 
As $\Delta U^*$ increases, the diameter of the satellite droplet increases 
and that of the main droplet decreases.

%------------------------------------------------------------------------------
% Table 1
%------------------------------------------------------------------------------
\begin{table}[!t]
\caption{Droplet size at $ka=0.7$ for $\Delta U^{*}=0$, 5.9, and 11.9}
\label{dsize_eachu}
%\begin{ruledtabular}
\begin{tabular}{ccc}
\hline \hline
  velocity difference $\Delta U^{*}$ \quad 
  & \quad main droplet $d_m/d$ \quad 
  & \quad satellite droplet $d_s/d$ \\ \hline
  0  & 1.87 & 0.318 \\
  5.9 & 1.858 & 0.4922 \\
  11.9 & 1.761 & 0.8662 \\ \hline \hline
\end{tabular}
%\end{ruledtabular}
\end{table}
%------------------------------------------------------------------------------

\begin{figure}[!t]
\begin{minipage}{0.325\linewidth}
\includegraphics[trim=2mm 0mm -3mm 1mm, clip, width=50mm, angle=-90]{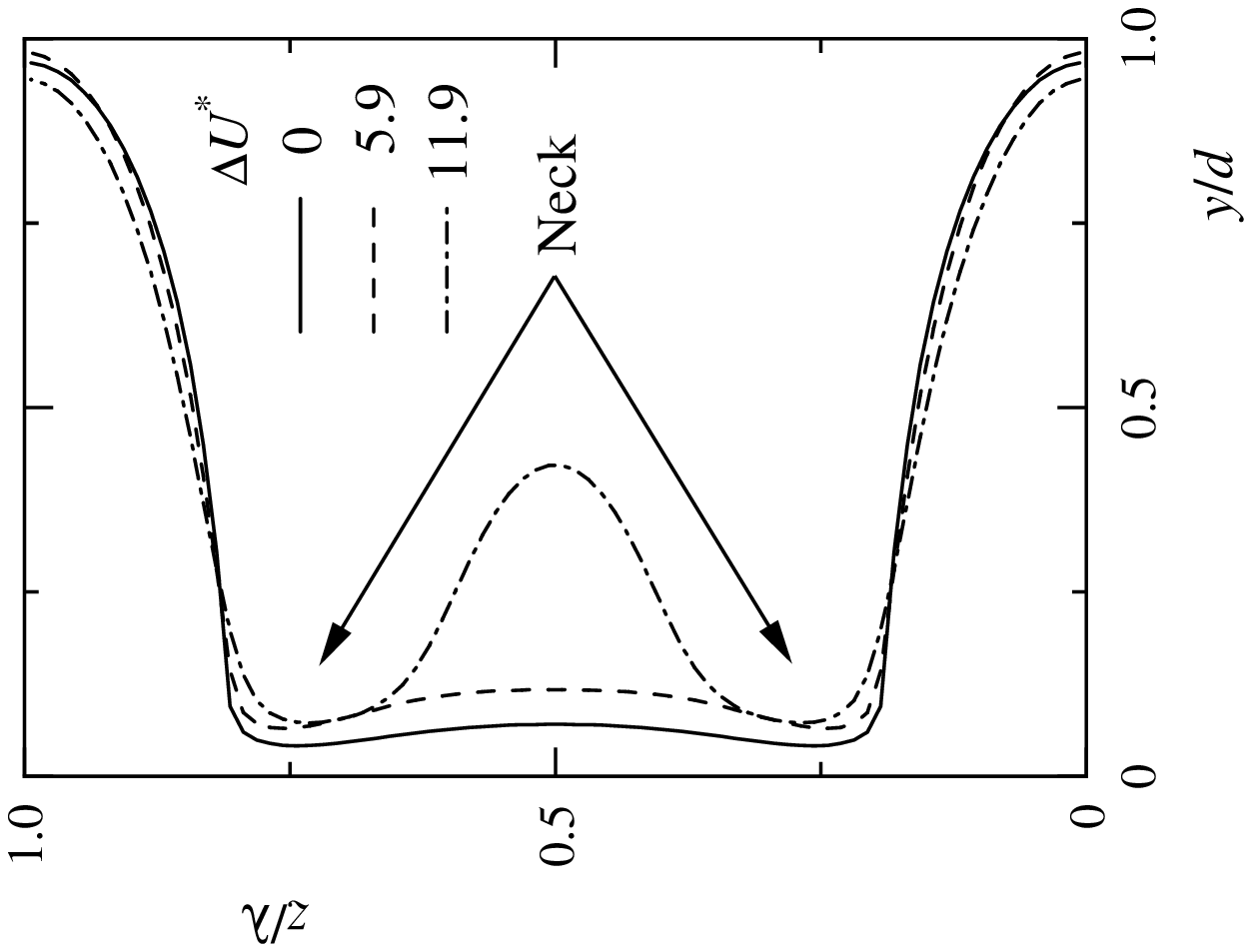} \\
(a) interface distributions
\end{minipage}
\begin{minipage}{0.325\linewidth}
\includegraphics[trim=2mm 0mm -3mm 1mm, clip, width=50mm, angle=-90]{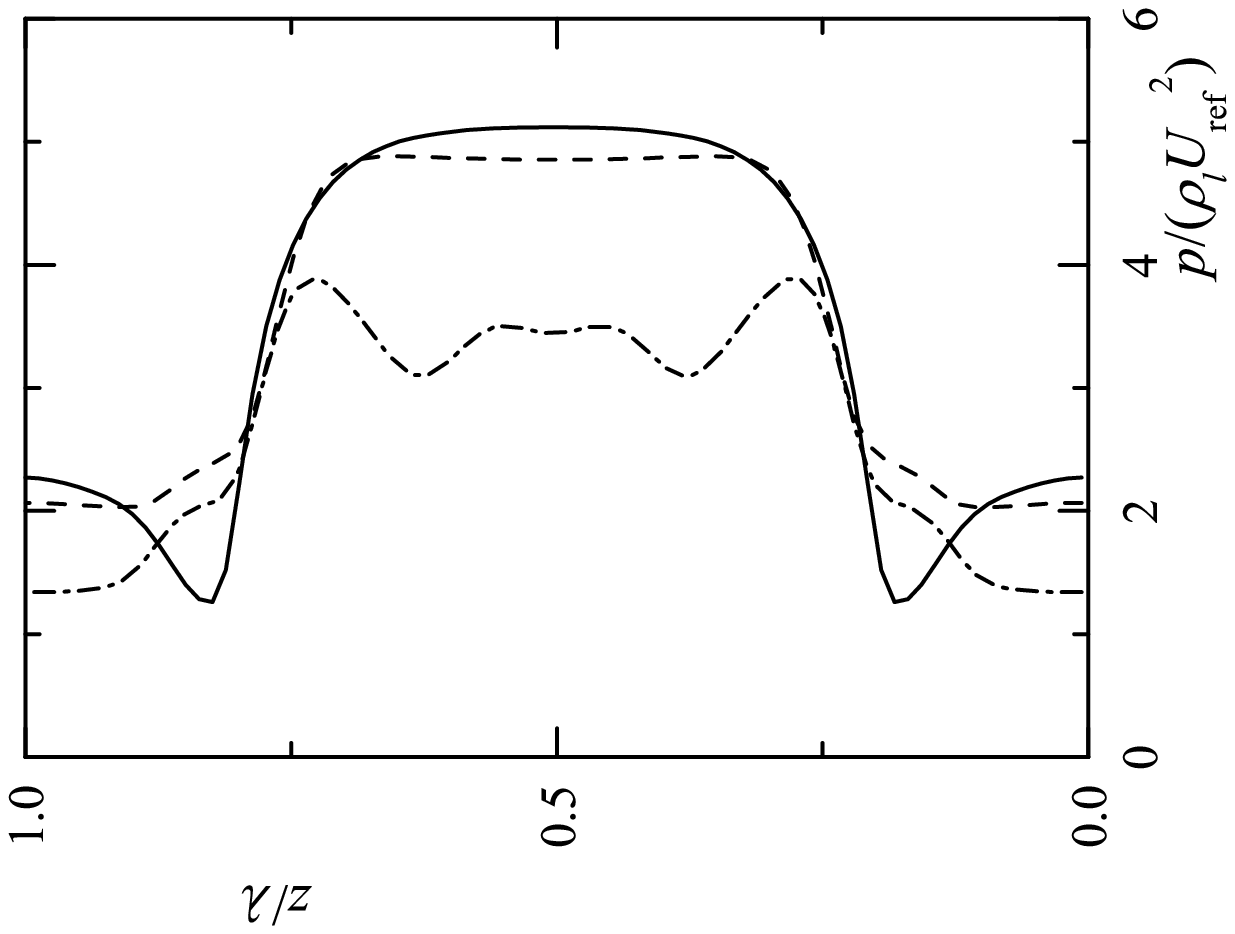} \\
(b) pressure distributions
\end{minipage}
\begin{minipage}{0.325\linewidth}
\includegraphics[trim=2mm 0mm -3mm 1mm, clip, width=50mm, angle=-90]{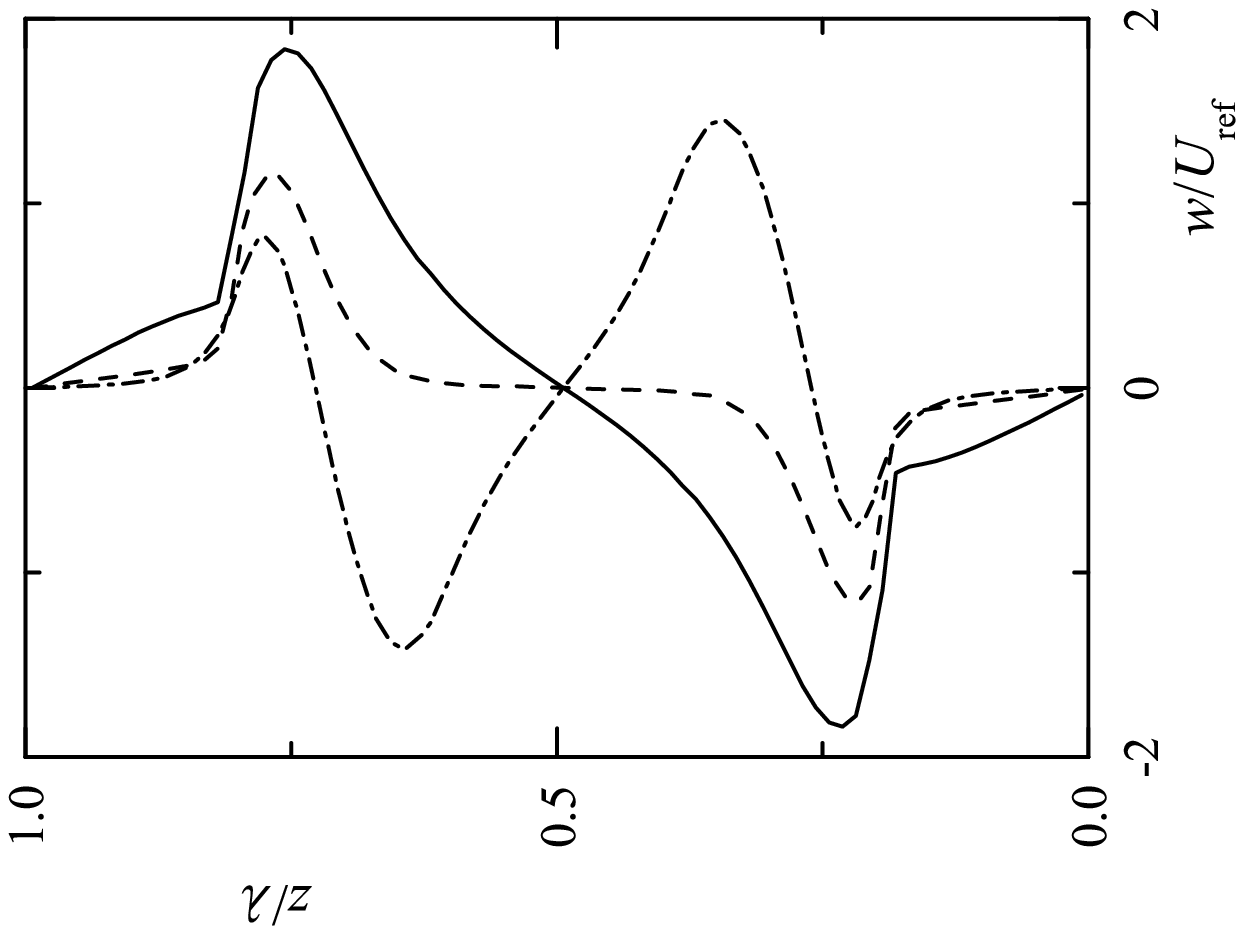} \\
(c) axial velocities
\end{minipage}
\caption{Time variations of ligament interface, pressure, and axial velocity 
distributions}
\label{neck_eachu}
\end{figure}

Figure \ref{neck_eachu} shows the distributions of the interface, 
pressure, and axial velocity of the ligament at $\Delta U^*=0$, 5.9, and 11.9 
for $ka=0.7$. 
Because the cross-section of the ligament affected 
by the shear of the airflow is non-circular, the interface distribution 
for the ligament at $\Delta U^*=5.9$ and 11.9 shows the equivalent radius 
of the circle obtained from the cross-sectional area of the ligament. 
The pressure is the distribution at $x/d=0$ and $y/d=0$ 
and the axial velocity is the average cross-sectional value in the ligament. 
The dimensionless time is $T=6.5$, 5.0, and 3.6 for $\Delta U^*=0$, 
5.9, and 11.9, respectively, 
and this figure shows the result immediately before the ligament breaks up. 
Regarding the interface distribution of the ligament, as $\Delta U^*$ increases, 
the radius near the center of the ligament increases, 
and the constriction forms at an earlier time. 
Regarding the pressure distribution, the pressure near the constriction 
is the highest at $\Delta U^*=5.9$ and 11.9. 
As a result, because the liquid in the center of the ligament does not move 
to the end, the flow velocity in the center of the ligament is almost zero 
at $\Delta U^*=5.9$. 
At $\Delta U^*=11.9$, the pressure difference between the constricted part 
and the central part increases, causing flow from the vicinity 
of the constricted part to the central part, as shown in the velocity distribution. 
Based on the existence of this liquid flow, it is considered that the diameter 
of the satellite droplet increases at $\Delta U^*=11.9$. 
Therefore, the diameter of the satellite droplet increases 
because the constrictions form early at both ends of the ligament interface.

%------------------------------------------------------------------------------
% Figure 13
%------------------------------------------------------------------------------
%------------------------------------------------------------------------------

%------------------------------------------------------------------------------
% Figure 14
%------------------------------------------------------------------------------
\begin{figure}[!t]
\begin{minipage}{0.48\linewidth}
\includegraphics[trim=0mm 0mm 0mm 0mm, clip, width=70mm]{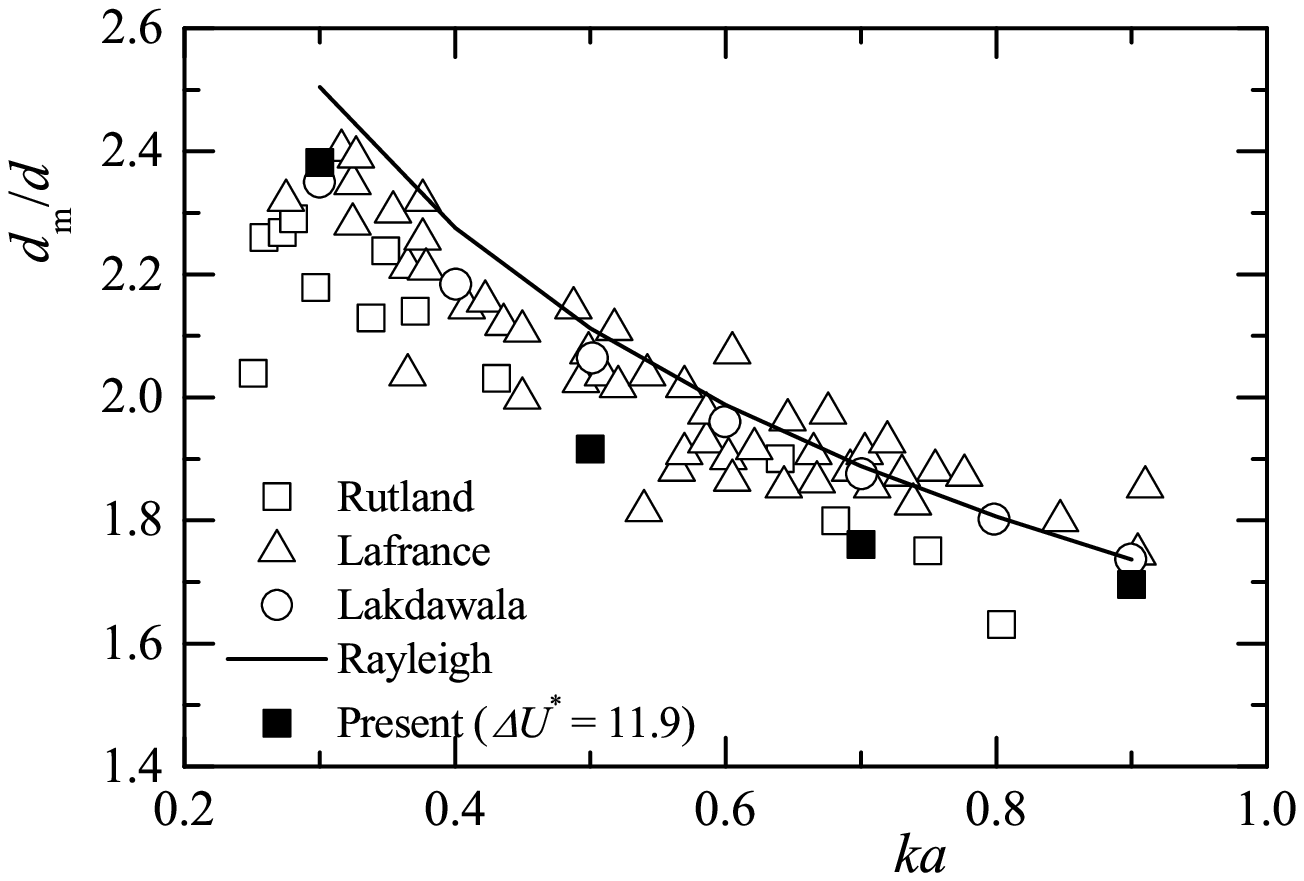} \\
(a) main drop
\end{minipage}
\hspace{0.02\linewidth}
\begin{minipage}{0.48\linewidth}
\includegraphics[trim=0mm 0mm 0mm 0mm, clip, width=70mm]{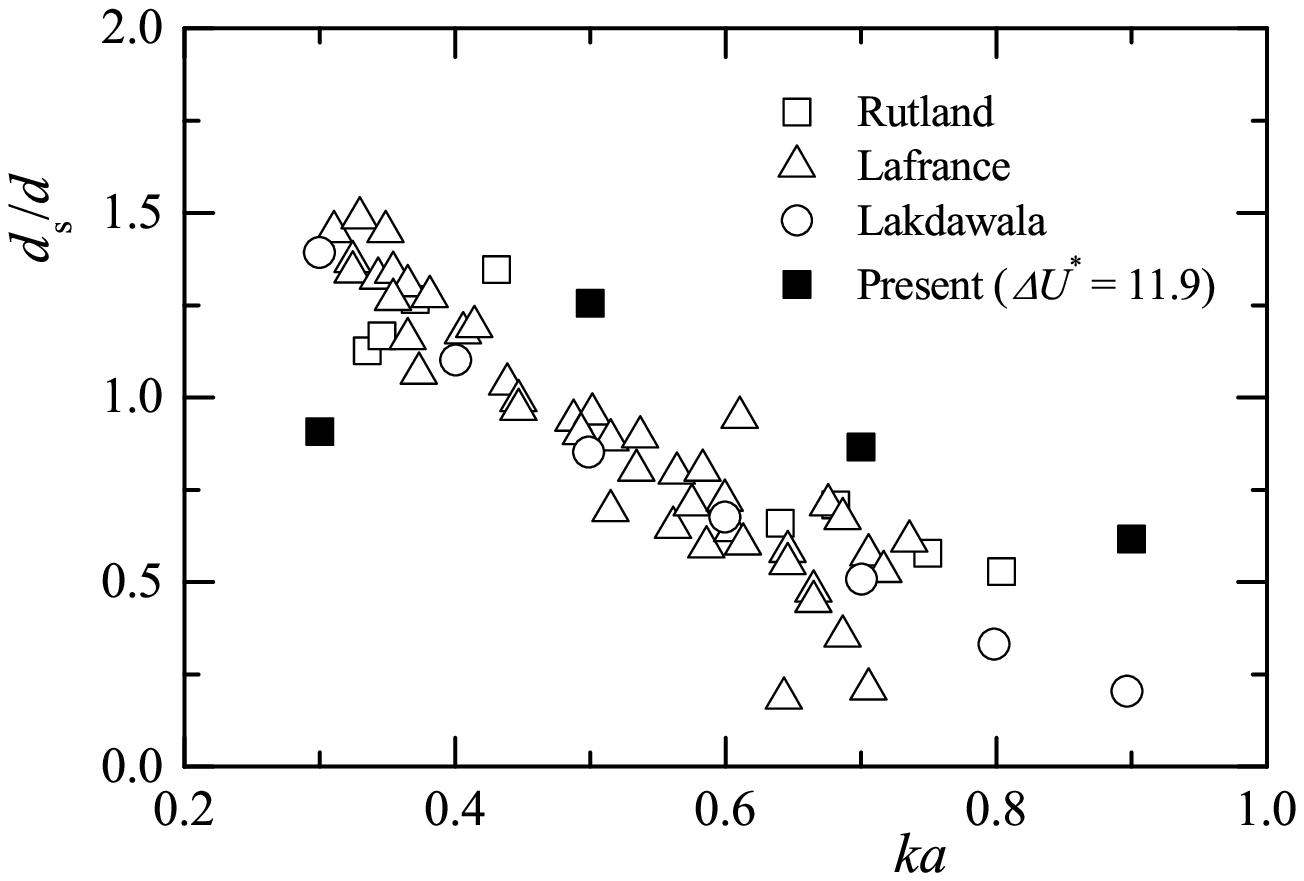} \\
(b) satellite drop
\end{minipage}
\caption{Comparison of droplet diameter with previous studies 
for various wavenumbers}
\label{dsize}
\end{figure}
%------------------------------------------------------------------------------

%------------------------------------------------------------------------------
% Figure 15
%------------------------------------------------------------------------------
\begin{figure}[!t]
\includegraphics[trim=0mm 0mm 0mm 0mm, clip, width=80mm]{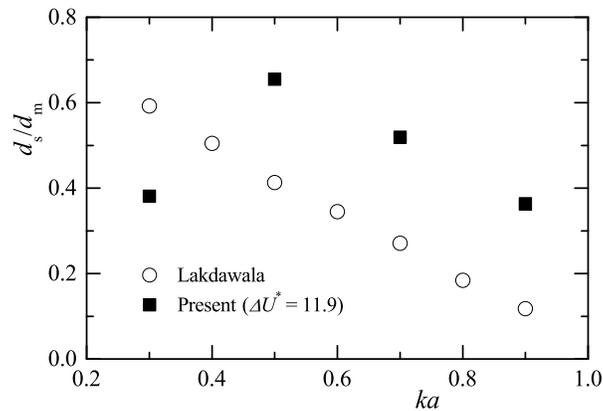}
\vspace*{-0.5\baselineskip}
\caption{Comparison of drop diameter ratio with previous study 
for various wavenumbers}
\label{dsdm}
\end{figure}
%------------------------------------------------------------------------------
Figure \ref{dsize} compares the droplet diameters obtained 
in this analysis with the results obtained from previous experiments 
on slow jets \cite{Rutland&Jameson_1971, Lafrance_1975} 
and a numerical analysis \cite{Lakdawala_et_al_2015}. 
Figure \ref{dsize}(a) shows the diameter of the main droplet 
and Fig. \ref{dsize}(b) shows the diameter of the satellite droplet. 
In Fig. \ref{dsize}(a), the solid line shows the diameter of the droplet 
calculated using linear theory \cite{Rayleigh_1878}. 
In this analysis, the diameter of the main droplet formed at $\Delta U^*=11.9$ 
is smaller than that observed in previous research. 
The relationship between the droplet diameter and the wavenumber 
is consistent with previous results. 
In Fig. \ref{dsize}(b), the diameter of the satellite droplet formed 
at $\Delta U^*=11.9$ is larger than that observed in previous studies, 
except at $ka=0.3$. 
At $ka=0.3$, the diameter of the satellite droplet is smaller than that at $ka=0.5$
because the diameter of the main droplet is larger than 
that for other wavenumbers.

Finally, Fig. \ref{dsdm} compares the ratio $d_s/d_m$ 
(i.e., the ratio of the diameter of the satellite droplet to that of the main droplet) 
with results obtained in a numerical analysis 
for slow jets \cite{Lakdawala_et_al_2015}. 
In this analysis, $d_s/d_m$ at $\Delta U^*=11.9$ is larger than 
that reported in a previous study except at $ka=0.3$. 
The diameters of the main droplet and satellite droplet 
are close to each other. 
This indicates that the ligament breaks up under the influence of 
the shear of the airflow, resulting in a decrease in the dispersion of 
the droplet diameter and homogenization of the diameter.

%##############################################################################
\section{Conclusion}
%##############################################################################

In this study, we performed a numerical analysis of the instability 
of a ligament in shear flow 
and investigated the effects of gas-liquid shear on the growth rate 
of the ligament interface, breakup time, and droplet diameter 
formed by the breakup. 
The following findings were obtained.

The ligament is stretched in the flow direction by the shear of the airflow. 
As the influence of the shear flow increases, 
the ligament deforms into a liquid sheet and a perforation forms 
at the center of the liquid sheet. 
The liquid sheet breaks up due to the growth of the perforation 
and contracts under the influence of surface tension, 
forming two ligaments with diameters smaller than that of the original ligament.

The shear of the airflow causes the original ligament to elongate, 
and the cross-section of the ligament becomes elliptical, 
which increases the instability. 
As a result, the growth rate of the ligament exceeds 
the theoretical value, 
and increases as the wavenumber of the initial disturbance increases. 
Therefore, the diameter of the formed droplet in shear flow 
decreases due to the increase in the wavenumber 
that governs the breakup of the ligament, 
and because the growth rate increases, the breakup time for the ligament 
becomes shorter.

As the velocity difference in the shear flow increases, 
constrictions of the ligament form earlier, 
and the diameter of the satellite droplet increases. 
As the diameter of the satellite droplet increases 
and the diameter of the main droplet decreases, 
the variation in the droplet diameter decreases, 
and the diameter becomes uniform.

%##############################################################################
% Acknowledgments
%##############################################################################
% If you have acknowledgments, this puts in the proper section head.
\begin{acknowledgments}
The numerical results in this research were obtained 
using supercomputing resources at the Cyberscience Center, Tohoku University. 
This research did not receive any specific grant from funding agencies 
in the public, commercial, or not-for-profit sectors. 
We would like to express our gratitude to Associate Professor Yosuke Suenaga 
of Iwate University for his support of our laboratory. 
The authors wish to acknowledge the time and effort of everyone involved 
in this study.
\end{acknowledgments}

%##############################################################################
\section*{Author Declarations}
%##############################################################################

\noindent
{\bf Conflicts of Interest}: The authors have no conflicts to disclose.

\noindent
{\bf Author Contributions}: 
H. Y. conceived and planned the research, 
and developed the calculation method and numerical codes. 
K. N. performed the simulations. 
H. Y. and K. N. contributed equally to analyzing data, reaching conclusions, 
and writing the paper.

%\noindent
%{\bf Author ORCID}: H. Yanaoka https://orcid.org/0000-0002-4875-8174

%##############################################################################
% Reference
%##############################################################################
% Create the reference section using BibTeX:
\bibliography{pof2022_nakayama}

%\clearpage
%##############################################################################
% Caption
%##############################################################################
% Show figure captions and table captions
%\listoffigures
%\listoftables

\end{document}